\documentclass[
  reprint,
  preprintnumbers,
  superscriptaddress,
  nofootinbib,
  amsmath,
  amssymb,
  aps,
  prd,
  floatfix
]{revtex4-2}

\pdfoutput=1

\usepackage{graphicx}
\usepackage{amsmath,amssymb,amsfonts,amsbsy,amstext,amsthm}
\usepackage{mathtools}
\usepackage{accents}
\usepackage{xcolor}
\usepackage{tabularx}
\usepackage{tensor}
\usepackage[
  colorlinks=true,
  urlcolor=blue,
  linkcolor=blue,
  citecolor=blue
]{hyperref}

\newcommand{\be}{\begin{equation}}
\newcommand{\ee}{\end{equation}}
\newcommand{\bea}{\begin{eqnarray}}
\newcommand{\eea}{\end{eqnarray}}

\newcommand{\tetrad}{\theta}
\newcommand{\cotetrad}{e}

\newcommand{\spinconnection}{\omega}

\newcommand{\Lorentz}{\Lambda}

\newcommand{\lapse}{\alpha}
\newcommand{\shift}{\beta}
\newcommand{\inducedmetric}{\gamma}
\newcommand{\normalvector}{\xi}
\newcommand{\momenta}{\pi}

\newcommand{\ftmc}{Center for Physical Sciences and Technology (FTMC), Saul\.{e}tekio av. 3, 10257 Vilnius, Lithuania}
\newcommand{\tartu}{Laboratory of Theoretical Physics, Institute of Physics, University of Tartu, W. Ostwaldi 1, 50411 Tartu, Estonia}
\begin{document}

\title{Hyperbolicity analysis of the linearised 3+1 formulation in the Teleparallel Equivalent of General Relativity}

\author{Cheng Cheng}
\email{cheng.cheng@ftmc.lt}
\affiliation{\ftmc}

\author{Mar\'ia-Jos\'e Guzm\'an}
\email{mjguzman@ut.ee}
\affiliation{\tartu}

\begin{abstract}
We study the properties of the principal symbol of the 3+1 equations of motion in Teleparallel Equivalent of General Relativity (TEGR) and assess the conditions for hyperbolicity. We use the Hamiltonian formulation based on the vectorial, antisymmetric, symmetric trace-free, and trace (VAST) decomposition of the canonical variables in the Hamiltonian formalism, and the Hamilton's equations previously presented in the literature. We study the system of differential equations at the linear level in one dimension, and show that the principal symbol has a sector with imaginary eigenvalues, which renders the system not hyperbolic. This situation is circumvented by identifying the problematic sectors, which are an isolated system and can be removed by a gauge fixing. We prove that the remaining system of equations is strongly hyperbolic. We also present the system in three dimensions. This is the first practical use of Hamilton's equations in TEGR, and our work can be extended for proving well-posedness in spherical symmetry, and establish numerical relativity setups in TEGR.

\end{abstract}

\maketitle

\section{Introduction}
\label{sec:intro}
Recent developments in gravitational wave (GW) physics \cite{LIGOScientific:2016aoc,LIGOScientific:2017vwq} stress the need for advanced numerical simulation methods with new mathematical formulations that can reframe numerical relativity while offering distinct computational advantages \cite{LIGOScientific:2016aoc}. The outstanding detection of a binary neutron star merger through GW170817, together with its electromagnetic counterpart GRB170817A \cite{LIGOScientific:2017vwq, Goldstein:2017mmi}, has further motivated the exploration of alternative, potentially more efficient approaches to simulations in the strong-gravity regime.

The exploration of alternative geometric frameworks for gravitational theories dates back to the inception of general relativity itself. In his later years, Einstein pursued the unification of gravitation and electromagnetism through the so-called teleparallel gravity \cite{Unzicker:2005in}. This framework introduces frames, or tetrads, as fundamental variables that mediate gravitational interactions. Tetrads have 16 independent components, and Einstein initially speculated that six of these could describe the electromagnetic field within a unified theory. He later recognised that these additional components are associated with Lorentz transformations that leave the metric invariant. Although the idea of teleparallelism was largely forgotten for several decades, research in the 1960s \cite{RevModPhys.48.393, Hayashi:1967se, Hayashi:1979qx} revisited the theory, considering it a potential solution to the energy-momentum complex problem. A renewed interest emerged in the 1990s, and since then, teleparallel gravity has gained traction both as an equivalent formulation of gravity from a different perspective and as a foundation for modified gravity theories.

The most general geometric framework for formulating gravity treats the metric and affine connection as independent fields. While the metric determines how distances are measured on a manifold, the affine connection prescribes the rules for parallel transport along curves. In general relativity (GR), the action is constructed from the Ricci scalar, which depends on the Levi-Civita connection--one that is both metric-compatible and torsion-free. This choice, however, is not obligatory. It is less commonly known that GR can be reformulated using connections that admit torsion or nonmetricity. In this work, we focus on a formulation in which the connection possesses torsion but exhibits vanishing curvature and nonmetricity. Specifically, the connection $\Gamma^\rho_{\mu\nu}$ is antisymmetric in its last two indices and is metric-compatible, meaning the associated covariant derivative satisfies $\nabla_{\rho}g_{\mu\nu} = 0$. This constitutes the so-called metric teleparallel framework, characterised by a flat manifold (i.e., vanishing curvature tensor) in which vectors can be parallel transported globally across the manifold. The primary geometric object is the torsion tensor, from which a torsion scalar $\mathbb{T}$ can be constructed. This scalar differs from the Ricci scalar $\mathbb{R}$ by a boundary term, resulting in the Teleparallel Equivalent of General Relativity (TEGR). With this scalar inserted into the action, the boundary term can be integrated out to yield Einstein’s equations. Consequently, all classical solutions of GR--such as cosmological and black hole spacetimes--are preserved in TEGR, and both theories possess the same number of degrees of freedom.

It is well established that GR can also be formulated in terms of tetrads, an approach that was explored in the context of numerical relativity several years ago. A formulation of general relativity using tetrads was developed in \cite{Buchman:2003sq}, building upon and confirming earlier results by Estabrook, Robinson, and Wahlquist \cite{Estabrook:1996wa}. These studies demonstrated that the tetrad equations for vacuum gravity can be expressed in an explicitly causal and symmetric hyperbolic form, independent of any particular time slicing, gauge choice, or coordinate specialisation. Tetrads have also been employed in numerical simulations of vacuum spacetimes extending across event horizons. Results from such simulations provided support for the Belinski-Khalatnikov-Lifshitz (BKL) conjecture, which posits that generic singularities are local, spacelike, and oscillatory \cite{Garfinkle:2003bb}. Despite these developments, the use of tetrads in numerical relativity has not seen widespread adoption. This may be due to the absence of troublesome physical problems in numerical relativity requiring this formalism\footnote{Private communication, Luis Lehner.} or the dominance of more established metric-based approaches. Nonetheless, renewed interest in tetrad formulations has emerged, particularly in relation to the Einstein–Dirac equations in the study of Dirac stars \cite{Alcubierre:2025wgj}. This is motivated by the fact that fermionic matter couples to gravity through tetrads, rendering them essential for incorporating spinor fields into gravitational theories. Extending these investigations to the broader metric-affine framework, of which teleparallel gravity constitutes a specific case, would be of significant interest.

In recent years, there has been growing interest in developing numerical relativity within the framework of teleparallel gravity, particularly the Teleparallel Equivalent of General Relativity (TEGR). Several studies have suggested that constructing a 3+1 formalism in both metric and symmetric teleparallel gravity may offer computational advantages for numerical simulations \cite{Capozziello:2021pcg, Peshkov:2022cbi, Olivares:2021cfa}. Within the context of TEGR, notable progress has been made, including a 3+1 decomposition of the Lagrangian equations of motion employing a tetrad split aligned with the Arnowitt-Deser-Misner (ADM) decomposition of the metric \cite{Capozziello:2021pcg}, a first-order formulation amenable to numerical implementation \cite{Olivares:2021cfa, Peshkov:2022cbi}, and a Hamiltonian 3+1 decomposition of TEGR \cite{Pati:2022nwi}. While the Hamiltonian formalism for TEGR has been extensively studied in earlier works \cite{Ferraro:2016wht, Blixt:2020ekl}, the covariant approach, generalizing the Weitzenböck gauge, has a more recent origin \cite{Golovnev:2021omn}. A less explored yet equally important framework is Symmetric Teleparallel Equivalent of General Relativity (STEGR), for which initial steps toward a 3+1 decomposition have been taken in \cite{Guzman:2023oyl}. It is argued that, despite the dynamical equivalence of the field equations, differences can emerge in their 3+1 forms and in the Hamilton's equation for $\dot{\pi}^{ij}$. This arises from the freedom in the boundary terms after decomposing the Lagrangian, leading to structural changes in the equations of motion in the functional dependence of the lapse and shift. An ADM formulation of STEGR in the coincident gauge has also been presented in \cite{DAmbrosio:2020nqu}; however, since the starting point is a 3+1 decomposition of the Einstein–Hilbert action, it does not exhibit significant deviations from standard GR. These foundational developments set the stage for future work in numerical relativity within modified teleparallel theories, such as $f(T)$, $f(Q)$, and related extensions. Given the current underdevelopment of their mathematical frameworks, the progress achieved thus far provides a critical foundation for further advances.

A crucial requirement for any 3+1 decomposition intended for implementation as a stable numerical code capable of making physical predictions is strong hyperbolicity. For systems of first-order partial differential equations, strong hyperbolicity is defined by the condition that the principal symbol (the matrix associated with spatial derivatives) possesses real eigenvalues and a complete set of eigenvectors. If the eigenvectors are incomplete, the system is classified as weakly hyperbolic. Strong hyperbolicity guarantees that the system admits a well-posed initial value problem. It is well known that the original ADM formulation of general relativity is only weakly hyperbolic. However, strong hyperbolicity can be achieved through suitable variable redefinitions, as in the BSSN formulation \cite{Baumgarte:1998te, Shibata:1995we}, or by introducing additional constraints, as in the CCZ4 system \cite{Dumbser:2017okk}. In this work, we investigate the hyperbolicity properties of the 3+1 equations of motion in TEGR. Due to the increased complexity of the equations--arising from the additional components of the tetrad--we analyse a simplified version of the system by linearising the dynamical variables. This toy model, introduced by Alcubierre in~\cite{article}, was originally used to demonstrate the weak hyperbolicity of the ADM equations and to show that strong hyperbolicity can be recovered through appropriate constraint additions. We will adopt this method to the TEGR framework.

This paper is organized as follows. In Section \ref{sec:tel}, we introduce the mathematical framework of the Teleparallel Equivalent of General Relativity (TEGR). Section \ref{sec:3p1} presents and compares the 3+1 decompositions in both Lagrangian and Hamiltonian formulations, and sets up Hamilton’s equations of TEGR for subsequent analysis. In Section \ref{sec:Lin}, we derive the linearised form of the evolution equations and constraints, while Section \ref{sec:Hyperb} is devoted to analysing their hyperbolicity properties. We conclude with a summary and outlook in Section \ref{sec:conc}, and present additional information in the two Appendices \ref{p1d} and \ref{principal3d}.

\section{Teleparallel gravity and 3+1 decomposition}
\label{sec:tel}
In this section, we introduce the underlying mathematical concepts of the metric teleparallel framework, and we present the TEGR Lagrangian and the equations of motion derived from it. We then establish the foundations of the most general 3+1 decomposition of the tetrad field, which we employ in Section \ref{sec:3p1}.

\subsection{Theoretical Framework}
We adopt the mostly positive sign convention for the Minkowski metric, i.e., $\eta_{AB} = \mathrm{diag}(-1, 1, 1, 1)$. Greek letters $\mu, \nu, \rho,...$ denote spacetime indices, while lowercase Latin letters $i,j,k,...= 1, 2, 3,$ are reserved for spatial indices. Lorentz tangent space indices are denoted by uppercase Latin letters $A, B, C,...$, with their spatial components indicated by lowercase Latin letters $a, b, c,...= 1, 2, 3$. We consider the tetrad fields defined at each point in spacetime, with components $\tetrad\indices{^A_\mu}$ and the inverse components $\cotetrad\indices{^\mu_A}$, which define the spacetime metric as follows
\begin{equation}
    g_{\mu\nu} = \eta_{AB} \tetrad^A{}_{\mu} \tetrad^B{}_{\nu}, \hspace{4mm}   \eta_{AB} = g_{\mu\nu} \cotetrad^{\mu}{}_A \cotetrad^{\nu}{}_{B}.
    \label{eq:fundrel}
\end{equation}
The tetrad and co-tetrad components also satisfy orthonormality relations
\begin{equation}
\tetrad^A{}_{\mu}  \cotetrad^{\mu}{}_B = \delta^{A}_{B}, \hspace{4mm} \tetrad^A{}_{\mu}  \cotetrad^{\nu}{}_A = \delta^{\nu}_{\mu}.
\label{orthonorm}
\end{equation}
Lorentz indices can be converted to spacetime indices and vice versa through contraction with the tetrad or cotetrad components. Specifically, a spacetime index, superscripted $\mu$, is mapped to a Lorentz index, superscripted $A$, via contraction with a tetrad $\theta^A{}_\mu$, while a covariant spacetime index, subscripted $\mu$, is mapped to a Lorentz index, subscripted $A$, using the inverse tetrad $e_A{}^\mu$. Lorentz indices are raised and lowered using the Minkowski metric $\eta_{AB}$, whereas spacetime indices are raised and lowered with the spacetime metric $g_{\mu\nu}$.

Building upon this structure, we introduce a curvature-free, metric-compatible spin connection $\spinconnection\indices{^A_B_\mu}$, whose components are defined by
\begin{equation}
\label{eq:spinconnlor}
    \spinconnection\indices{^A_B_\mu} = -(\Lorentz^{-1})\indices{^C_B} \partial_\mu \Lorentz\indices{_C^A}, 
\end{equation}
where $\Lorentz\indices{_C^A}$ are $4 \times 4$ matrix elements of the Lorentz group. The spin connection enters the teleparallel action and, thus, in this formulation the Lorentz matrices are treated as canonical fields \cite{Golovnev:2021omn, Golovnev:2017dox}.

The main building block in teleparallel theories of gravity is the torsion tensor, which depends on both the tetrad and the spin connection as
\begin{equation}
    T\indices{^\rho_\mu_\nu}=e\indices{_A^\rho}(\partial_\mu\theta\indices{^A_\nu}-\partial_\nu\theta\indices{^A_\mu}+\omega\indices{^A_B_\mu}\theta\indices{^B_\nu}-\omega\indices{^A_B_\nu}\theta\indices{^B_\mu}).\label{eq.5}
\end{equation}
From it, we build the torsion scalar $\mathbb{T}$

\begin{align}
    \mathbb{T}=\frac{1}{4}T\indices{^\rho_\mu_\nu}T\indices{_\rho^\mu^\nu}+\frac{1}{2}T\indices{^\rho_\mu_\nu}T\indices{^\mu^\nu_\rho} -T\indices{^\rho_\mu_\rho}T\indices{^\sigma_\mu_\sigma}, 
\label{Tscalar1}
\end{align}
which is related to the Ricci scalar $\mathbb{R}$ of general relativity by a total divergence
\begin{equation}
\mathbb{R} = - \mathbb T + 2 \tetrad \partial_{\mu} (\cotetrad T^{\mu} ),\label{EqRTB}
\end{equation}
where $T^{\mu} = T\indices{^\alpha_\alpha^\mu}$, $\tetrad = \det(\tetrad\indices{^A_\mu} )$ is the determinant of the tetrad, and $e$ is the determinant of the co-tetrad. 
Alternatively, the torsion scalar can be written as
\begin{align}
\mathbb T = T^{\rho\mu\nu}S_{\rho\mu\nu},
\label{Tscalar2}
\end{align}
where $S_{\rho\mu\nu}$ is the so-called superpotential, defined in terms of the torsion tensor as
\begin{align}
    S_{\rho\mu\nu}=\frac{1}{2}T_{\rho\mu\nu}+T_{[\mu\nu]\rho}+2g_{\rho[\mu}T^\sigma{}_{\nu]\sigma}.
\label{superp}
\end{align} 
We use the torsion scalar to define the TEGR Lagrangian $\mathcal{L}_{\mathrm{TEGR}}=-\tetrad \mathbb T/2\kappa$, where the constant $\kappa=8\pi G/c^{4}$ is composed of the gravitational constant $G$ and the speed of light $c$. Substituting the relation between $\mathbb R$ and $\mathbb T$ in \eqref{EqRTB} into the Einstein-Hilbert action gives the TEGR action,
\begin{align}
    S_\mathrm{TEGR}=-\frac{1}{2\kappa}\int \mathrm{d}^4 x \tetrad \mathbb{T}.\label{Stegr}
\end{align}
Since the boundary term in Eq.~\eqref{EqRTB} only contributes to a total divergence, it can be integrated out under the four-dimensional integral in the action. As a result, TEGR yields the same equations of motion as GR, sharing its classical solutions and number of degrees of freedom. The field equations are obtained by varying the TEGR action with respect to the tetrad $\tetrad^{A}{}_{\nu}$, and in vacuum they take the form
\begin{align}
    \cotetrad\indices{^\nu_A}\mathbb T-2e\partial_{\lambda}(\tetrad\cotetrad\indices{^\sigma_A}S\indices{_\sigma^\lambda^\nu}) - 2 \cotetrad\indices{^\mu_A} T\indices{^\rho_\sigma_\mu}S\indices{_\rho^\sigma^\nu} = 0. 
    \label{eomTEGR}
\end{align}
These equations are dynamically equivalent to Einstein's field equations, confirming the classical equivalence between TEGR and GR.

The torsion scalar defined in Eq.~\eqref{Tscalar1} serves as the fundamental building block for a variety of modified theories of gravity within the teleparallel framework. One notable extension involves relaxing the fixed coefficients in front of the three quadratic torsion terms, leading to the class of theories known as ``new general relativity'' \cite{Hayashi:1979qx}. A more widely studied generalization is to promote the torsion scalar to an arbitrary function, giving rise to $f(\mathbb{T})$ gravity. This theory is conceptually analogous to $f(\mathbb R)$ gravity in the metric formulation, yet it exhibits fundamentally different physical properties and dynamics. Although we do not explore these extensions of TEGR in the present work, they present directions for future research. The theoretical foundations developed here intend to serve as a basis for such investigations.

In the remainder of this section, we study the dynamics of the equations of motion in Eq.~\eqref{eomTEGR} from a Hamiltonian perspective. In particular, we perform a 3+1 decomposition of the equations of motion, which is achieved by computing Hamilton's equations. While such a decomposition can also be carried out directly at the level of the Lagrangian equations of motion, they offer a dynamically equivalent result. The Hamiltonian formulation offers a deeper understanding of the underlying structure of TEGR, and it helps to visualise and manage the extra constraints coming from using the tetrad as the dynamical variables. This approach also naturally prepares the formalism for future applications to modified teleparallel theories discussed earlier.

\subsection{3+1 decomposition of the tetrad}
Our first step in deriving Hamilton's equations of motion for TEGR is performing a suitable 3+1 decomposition that accounts for the extra independent components that the tetrad field encodes. Unlike the metric, the tetrad field contains 16 independent components, reflecting that, for any given metric, there is an infinite family of tetrads related to each other by local Lorentz transformations. In this case, the tetrad 3+1 decomposition must reproduce the ADM decomposition of the metric without requiring any gauge fixing in the tetrad components.

We foliate the four-dimensional spacetime manifold described by the metric $g_{\mu\nu}$ into a family of three-dimensional hypersurfaces $\Sigma_t$, labeled by a global time coordinate $t$. Each hypersurface has an embedded three-dimensional induced metric $\inducedmetric_{ij}$. We also introduce the standard ADM variables, the lapse function $\alpha$ and the shift vector $\beta^{i}$, and the ADM metric:
\be
g_{00}=-\lapse^2+\shift^i\shift^j\inducedmetric_{ij}, \quad g_{0i}=\shift_i,\quad g_{ij}=\inducedmetric_{ij},
\label{eq:ADMmetric}
\ee
and its inverse
\be
g^{00}=-\frac{1}{\lapse^2},\quad g^{0i}=\frac{\shift^i}{\lapse^2},\quad g^{ij}=\inducedmetric^{ij}-\frac{\shift^i\shift^j}{\lapse^2}.
\label{eq:ADMinv}
\ee
We consider the spatial components $\theta^{A}{}_{i}$ of the tetrad as the canonical variables, instead of the induced metric. These objects are related to each other by virtue of
\be
\tetrad^A{}_i\tetrad^B{}_j\eta_{AB}=\inducedmetric_{ij}.
\ee
The spatial components $\theta^{A}{}_{i}$ of the tetrad refer to the split of spacetime indices $\mu$, instead of the Lorentz (tangent space) indices $A$. The latter will become relevant at a later stage, but is not required for the present discussion. To complete the decomposition, we must also introduce the temporal component of the tetrad $\theta^{A}{}_{0}$ consistent with the ADM decomposition given in Eq.~\eqref{eq:ADMmetric}. A convenient choice is \cite{Pati:2022nwi}
\be
\tetrad^A{}_0=\lapse\normalvector^A+\shift^i\tetrad^A{}_i,
\label{ADMtetrad}
\ee
with the vector field $\normalvector^{A}$ that satisfies the normalisation condition
\be
\eta_{AB}\normalvector^A\normalvector^B=\normalvector_A\normalvector^A=-1.
\label{normxi}
\ee
In addition, we must impose an orthonormality condition between the spatial part of the tetrad and the normal vector, which reads
\be 
\eta_{AB}\normalvector^B\tetrad^A{}_i=\normalvector_A\tetrad^A{}_i=0. 
\label{nvect}
\ee
We note that the vector $\normalvector^{A}$ that satisfies all these properties can be written as (see, for instance, \cite{Castellani:1981ue}),
\be 
\xi^{A} = -\dfrac16 \epsilon\indices{^A_B_C_D} \tetrad^{B}{}_{i} \tetrad^{C}{}_{j} \tetrad^{D}{}_{k} \epsilon^{ijk}.
\ee 
We show that this tetrad decomposition reproduces the ADM metric \eqref{eq:ADMmetric}, since
\begin{equation}
    \begin{split}
        g_{00} & = \eta_{AB} \tetrad^{A}{}_0 \tetrad^{B}{}_{0} \\
& = \eta_{AB}( \alpha \normalvector^{A} + \beta^{i}\tetrad^{A}{}_{i} ) (\alpha \normalvector^{B} + \beta^{j} \tetrad^{B}{}_{j} ) \\
& = - \alpha^2 + \beta_{i} \beta^{i},
    \end{split}
\end{equation}
and
\begin{equation}
    g_{0j} = \eta_{AB} \tetrad^{A}{}_0 \tetrad^{B}{}_j = \eta_{AB}(\alpha \normalvector^{A} + \beta^{i}\tetrad^{A}{}_i) \tetrad^{B}{}_j = \beta_j.
\end{equation}

We must also propose a suitable ADM-like split of the inverse tetrad $\cotetrad_{A}{}^{\mu}$ as follows
\be
\cotetrad_A{}^0=-\frac{1}{\alpha}\normalvector_A, \indent  \cotetrad_A{}^i=\tetrad_A{}^i+\normalvector_A\frac{\shift^i}{\lapse}.
\label{ADMinvE}
\ee
It is important to note that the right-hand side of the second equation in \eqref{ADMinvE} defines the object $\theta_{A}{}^{i}$, which is used as the shorthand for $\theta_{A}{}^{i} = \eta_{AB} \gamma^{ij} \tetrad^{B}{}_{j}$. In our main result, we will minimise the use of the $\theta_{A}{}^{i}$, in order to avoid confusion with the inverse tetrad components $e_{A}{}^{i}$.
This decomposition ensures consistency with the inverse metric components. Specifically, we recover:
\begin{equation}
    \begin{split}
    g^{00} & = \eta^{AB} \cotetrad_{A}{}^{0} \cotetrad_{B}{}^{0} \\
    & = \eta^{AB} \left(-\dfrac{1}{\alpha} \xi_{A} \right) \left(-\dfrac{1}{\alpha} \xi_{B} \right) = -\dfrac{1}{\alpha^2},
\end{split}
\end{equation}
\begin{equation}
    \begin{split}
         g^{0i} & = \eta^{AB} \cotetrad_{A}{}^{0} \cotetrad_{B}{}^{i} \\
    & = -\dfrac{1}{\alpha} \xi_{A} \left( \tetrad_{B}{}^{i} + \xi_{B} \dfrac{\beta^{i}}{\alpha}\right)\eta^{AB} = \dfrac{1}{\alpha^2} \beta^{i},
    \end{split}
\end{equation}
\begin{equation}
    \begin{split}
        g^{ij} & = \cotetrad_{A}{}^i \cotetrad_{B}{}^j \eta^{AB}\\
& = \eta_{AB} \left( \tetrad_{A}{}^i + \normalvector_A \dfrac{\beta^{i}}{\alpha} \right) \left( \tetrad_{B}{}^j + \normalvector_B \dfrac{\beta^{j}}{\alpha} \right) \\
& = \gamma^{ij} - \dfrac{\beta^i \beta^j}{\alpha^2}.
    \end{split}
\end{equation}   
Additionally, we provide a useful identity relating the spatial tetrad and cotetrad components:
\begin{align}
    \eta_{AD}\inducedmetric^{kl}\tetrad^C{}_k \tetrad^D{}_l = \tetrad^{C}{}_{k} \tetrad_{A}{}^{k} = \delta^C_A+\eta_{AD} \normalvector^C \normalvector^D.
\end{align}
We assume that the 3+1 tetrad decomposition  respects the conditions for a proper spacetime foliation, which have been recently discussed in \cite{Blixt:2024aej}. We observe that our 3+1 tetrad decomposition consists of 16 canonical variables $(\alpha,\beta^{i},\tetrad^{A}{}_{i})$, spanning the 16 independent components of the tetrad field. It is worth noting that the $3+1$ tetrad decomposition presented in \cite{Capozziello:2021pcg} was performed under a specific gauge choice that lacks generality. Although not relevant for TEGR, this assumption can become problematic for more general teleparallel theories such as $f(\mathbb T)$ gravity.
\section{3+1 formulations of TEGR in Lagrangian and Hamiltonian formalisms}
\label{sec:3p1}
In this section we present the 3+1 decomposition of the TEGR Lagrangian, and its reformulation within the Hamiltonian framework. Our goal is to present the derivation of Hamilton's equations, which correspond to the 3+1 decomposed Lagrangian equations expressed in terms of the canonical variables. 

Following the definitions and geometric constructions introduced in the previous section, we proceed by applying the 3+1 decomposition of the tetrad field to all relevant quantities appearing in the TEGR Lagrangian. We obtained
\begin{equation}\label{L3p1}
    \begin{split}      &\mathcal{L}_\mathrm{TEGR}=\mathcal{L}_S+\frac{\sqrt{\inducedmetric}}{2\lapse}M\indices{^i_A^j_B}T\indices{^A_0_i}T\indices{^B_0_j}\\& -\frac{\sqrt{\inducedmetric}}{\lapse}T\indices{^A_0_i}T\indices{^B_k_l}\left[M\indices{^i_A^l_B}\shift^k-\frac{\lapse}{\kappa}\inducedmetric^{il}\left(\frac{1}{2}\normalvector_B\tetrad_A{}^k-\normalvector_A\tetrad_B{}^k \right) \right],
    \end{split}
\end{equation}
where the time derivatives of the tetrad field are encoded in the components of $T^{A}{}_{0i}$, and $M\indices{^i_A^j_B}$ is defined as
\be
\begin{split}
    M\indices{^i_A^j_B}=\frac{1}{2\kappa}\left(\inducedmetric^{ij}\eta_{AB}+\normalvector_A\normalvector_B\inducedmetric^{ij}+\tetrad_A{}^j\tetrad_B{}^i-2\tetrad_A{}^i\tetrad_B{}^j\right),
\label{HessianTEGR}
\end{split}
\ee
and the term $\mathcal{L}_S$ depending only on spatial derivatives of the tetrad     
\begin{equation}
    \begin{split}
    \mathcal{L}_S&=\frac{\sqrt{\inducedmetric}}{\lapse}T^A{}_{ij}T^B{}_{kl}\shift^i\left[\frac{1}{2}M\indices{^j_A^l_B}\shift^k\right.\\&\left.-\frac{\lapse}{\kappa}\inducedmetric^{jl}\left(\frac{1}{2}\normalvector_B\tetrad_A{}^k-\normalvector_A\tetrad_B{}^k \right) \right] +\frac{\lapse\sqrt{\inducedmetric}}{2\kappa}{}^3\mathbb{T},
\label{SpatialL}
\end{split}
\end{equation}
where ${}^3\mathbb{T}$ is the spatial part of the torsion scalar and is defined as follows
\begin{equation}
\begin{split}        
{}^3\mathbb{T}&=H\indices{_A_B^i^j^k^l}T\indices{^A_i_j}T\indices{^B_k_l}\\& =\left(\frac{1}{2}\tetrad_B{}^{[i}\inducedmetric^{j][k}\tetrad_A{}^{l]}-\frac{1}{4}\eta_{AB} \inducedmetric^{k[i}\inducedmetric^{j]l}\right.\\&-\tetrad_A{}^{[i}\inducedmetric^{j] [k}\tetrad_B{}^{l]} \biggr)T\indices{^A_i_j}T\indices{^B_k_l}.
\label{eq.22}
\end{split}
\end{equation}
With this $3+1$ Lagrangian, we compute the canonical momenta associated to the canonical variables. These correspond to the 16 phase space functions $(\pi, \pi_i, \pi_{A}{}^{i})$, that is the momenta conjugate to lapse $\alpha$, shift $\beta^{i}$, and spatial components of the tetrad $\theta^{A}{}_{i}$, respectively. Since the only components in the Lagrangian that contain time derivatives of the spatial tetrad $\partial_0 \tetrad^{A}{}_{i}$ \footnote{We will use interchangeably the notation $\partial_0 A$ and $\dot{A}$ for time derivatives of fields.} appear through the torsion components $T^{A}{}_{0i}$, the canonical momenta can be easily obtained
\begin{equation}
\begin{split}
    \momenta_{A}{}^{i} &= \dfrac{\partial \mathcal{L}_\mathrm{TEGR}}{\partial_0  \tetrad^A{}_\mu} = \dfrac{\partial \mathcal{L}_\mathrm{TEGR}}{\partial T^{A}{}_{0i} }\\& = \dfrac{\sqrt{\gamma} }{ \alpha} \biggl[ M\indices{^i_A^l_B} (T^{B}{}_{0l} - T^{B}{}_{ml}\shift^m) \\&\left.+ \dfrac{\alpha}{\kappa}T^{B}{}_{ml} \inducedmetric^{il}\left(\dfrac12 \normalvector_{B}\tetrad_{A}{}^{m} - \normalvector_{A}\tetrad_{B}{}^{m} \right)  \right].
\end{split}
\label{piT}
\end{equation}

The conjugate momenta associated to the lapse and shift vanish since the Lagrangian does not contain time derivatives of those fields. These momenta lead to the primary constraints: \footnote{The symbol $\approx$ represents a weak equality, in the sense that the function vanishes when restricted to the constraint surface \cite{Sundermeyer:2014kha}. } 
\begin{equation}
\label{Calph}
    {}^\alpha C={}^\alpha \pi:=\frac{\partial \mathcal{L}_\mathrm{TEGR}}{\partial \partial_{0} \alpha}\approx 0,
\end{equation}
\begin{equation}
    \label{Cbeth}
     {}^\beta C_i={}^\beta \pi_i:=\frac{\partial \mathcal{L}_\mathrm{TEGR}}{\partial \partial_0 \beta^i}\approx 0. 
\end{equation}
We collectively denote these constraints as $C_{A}=({}^\alpha C, {}^\beta C_i )$, with corresponding Lagrange multipliers $\lambda^{A} = ({}^{\alpha}\lambda, {}^{\beta}\lambda^i )$ introduced in the primary Hamiltonian. In the literature \cite{Ferraro:2016wht,Blixt:2020ekl}, the canonical variables are chosen as $(\tetrad^{A}{}_{0}, \tetrad^{A}{}_{i})$ instead of $(\lapse,\shift^{i},\tetrad^{A}{}_{i})$. In such cases, the primary constraints appear as $\momenta_{A}{}^{0}\approx0$, reflecting the absence of time derivatives of $\tetrad^{A}{}_{0}$ in the Lagrangian. These primary constraints are generic features of teleparallel theories based on torsion and tetrad fields; therefore, they appear not only in TEGR but also in its extensions \cite{Ferraro:2018tpu,Ferraro:2018axk,Ferraro:2020tqk,Guzman:2020kgh}.

Additional primary constraints are obtained from specific combinations of the canonical momenta and torsion components. They are given by
\begin{equation}
    ^{\mathcal{V}}C^{i} = -\dfrac{\kappa}{\sqrt{\gamma}} \normalvector^{A} \momenta_{A}{}^{i} + T^{B}{}_{jk} \gamma^{ik} \gamma^{jl} \tetrad^{A}{}_{l} \eta_{AB} \approx 0
    \label{LorC1}
\end{equation}
and 
\begin{equation}
^{\mathcal{A}} C^{ij}=-\frac{\kappa}{2\sqrt{\gamma}}\theta\indices{^A_k}\momenta\indices{_A^[^i}\gamma^{j]k}-\frac{1}{2} \gamma^{ik} \gamma^{jl} T\indices{^B_k_l} \normalvector_{B} \approx 0.
\label{LorC2}
\end{equation}
These constraints are associated with local Lorentz transformations, which allow us to describe a specific metric with infinite many choices of tetrads. In addition to the momenta of the tetrads, we must also consider the conjugate momenta of the Lorentz matrices $\Lambda^{A}{}_{B}$, which enter the theory through the teleparallel spin connection defined in Eq.~\eqref{eq:spinconnlor}. Following the approach in \cite{Golovnev:2021omn}, the momenta $P^{B}{}_{A}$ have been taken as the variation with respect to $\partial_0 \Lambda^{A}{}_{B}$,
\begin{equation}
    P^B{}_A = \frac{\partial \mathcal{L}_\mathrm{TEGR}}{\partial_0 \Lambda^A{}_B}=\momenta_C{}^i \eta_{AD}\left(\Lambda^{-1} \right)_E{}^B\eta^{C[E}\tetrad^{D]}{}_i.
\label{MomType1}
\end{equation}
While no antisymmetry has been assumed for the Lorentz matrices $\Lambda^{A}{}_{B}$ themselves, their combination with the momenta $P^{B}{}_{A}$ in the form of a primary constraint restricts the number of independent components. These additional primary constraints are given by
\begin{align}
\label{LorC3}
C^{AB} = P^{[A}{}_D\eta^ {B]C}\Lorentz_C{}^D+\momenta_C{}^i\eta^{C[B}\tetrad^{A]}{}_i \approx 0,
\end{align}
and area added to the primary Hamiltonian through Lagrange multipliers $\lambda_{AB}$. Although both sets of constraints in Eq.~\eqref{LorC1}-\eqref{LorC2} and Eq.~\eqref{LorC3} can be interpreted as Lorentz transformations, they play different roles in teleparallel theories and have been described as Lorentz constraints of type I and type II, respectively. More details about this classification can be found in \cite{Blixt:2022rpl}. In the analysis detailed in \cite{Golovnev:2021omn}, these extra constraints have vanishing Poisson brackets with the remaining primary constraints. It is thus reasonable to postulate that they are first-class constraints.
\subsection{TEGR's Hamiltonian}
To derive the TEGR Hamiltonian, in \cite{Pati:2022nwi} it is introduced an irreducible decomposition of the velocities $\dot{\tetrad}^{A}{}_{i}$ and their canonical momenta $\pi_{A}{}^{i}$ under the rotation group $\mathcal{O}(3)$. We refer the readers to the original work for more details. As mentioned, the velocities are contained in the components of $T^{A}{}_{0i}$, and can be inverted in terms of the momenta as

\begin{equation}
    \begin{split}
    T^C{}_{0k}&=-\frac{\lapse}{\kappa}\left(M^{-1}\right)\indices{_i^A_k^C}T\indices{^B_m_l}\inducedmetric^{il}\left(\frac{1}{2}\normalvector_B \tetrad_A{}^m -\normalvector_A \tetrad_B{}^m \right)\\&+\left(M^{-1}\right)\indices{_i^A_k^C}\frac{\alpha}{\sqrt{\inducedmetric}}\pi_A{}^i+T\indices{^C_m_k}\shift^m,
    \end{split}
\end{equation}
where
\begin{equation}
\left(M^{-1}\right)\indices{_i^A_k^C}=\frac{\kappa}{2}\left(\inducedmetric_{ik}\inducedmetric^{mn}\tetrad^A{}_m \tetrad^C{}_n+\tetrad^A{}_k\tetrad^C{}_i-\tetrad^A{}_i \tetrad^C{}_k\right)
\end{equation}
is the Moore-Penrose pseudoinverse, and facilitates this inversion since it inverts the matrix $M\indices{^i_A^l_B}$ by blocks, as discussed in \cite{Ferraro:2016wht,Pati:2022nwi}. By performing all pertinent computations, the primary Hamiltonian for the covariant TEGR is as
\onecolumngrid
\begin{equation}
\begin{split}
    \mathcal{H}_\mathrm{TEGR}&=\lapse  \left[ \frac{\kappa}{2\sqrt{\inducedmetric}} {}^{\mathcal{S}}{\momenta}_{ij} {}^{\mathcal{S}} \momenta^{ij} -\frac{3\kappa}{4\sqrt{\inducedmetric}}{}^{\mathcal{T}} \momenta {}^{\mathcal{T}} \momenta -\frac{\sqrt{\inducedmetric}}{2\kappa}{}^3 \mathbb{T}-\normalvector^A\partial_i \momenta_A{}^i+\momenta_A{}^i\spinconnection^A{}_{Bi}\normalvector^B \right] \\
    &+\shift^j\left[ -\tetrad^A{}_j \partial_i \momenta_A{}^i+\momenta_A{}^i\spinconnection^A{}_{Ci} \tetrad^C{}_j-\momenta_A{}^i T^A{}_{ij} \right] -\lambda^{A}C_{A} -\lambda_{AB}\left( P^{[A}{}_D\eta^ {B]C}\Lorentz_C{}^D+\momenta_C{}^i\eta^{C[B}\tetrad^{A]}{}_i \right) \\
  &-{}^{\mathcal{V}}\lambda_i\left(\frac{{}^{\mathcal{V}} \pi^i\kappa}{\sqrt{\inducedmetric}}+T^B{}_{jk}\inducedmetric^{ik}\inducedmetric^{jl}\tetrad^A{}_l\eta_{AB}\right)-{}^{\mathcal{A}}\lambda_{ij}\left(\frac{{}^{\mathcal{A}} \pi^{ij}\kappa}{\sqrt{\inducedmetric}}-\dfrac12 \inducedmetric^{ik}\inducedmetric^{jl}T^B{}_{kl}\normalvector_B \right)+\partial_i \left(\momenta_A{}^i\tetrad^A{}_0 \right).
  \label{eq.31}
\end{split}
\end{equation}
\twocolumngrid
We note that the final term $\partial_i \left(\momenta_A{}^i\tetrad^A{}_0 \right)$ is a boundary term and contains spatial derivatives of the lapse and shift. In \cite{Pati:2022nwi}, this term was integrated out and therefore does not play a role in the derivation of Hamilton's equations. Alternatively, the Hamiltonian can be written in terms of the original canonical variables before the irreducible decomposition,
\onecolumngrid
\begin{equation}
\begin{split}
    \mathcal{H}_\mathrm{TEGR}&=\lapse  \left[\frac{ \kappa}{4\sqrt{\inducedmetric}}\left[\momenta_A{}^i\momenta_B{}^l\tetrad^A{}_k \tetrad^B{}_j\inducedmetric^{jk}\inducedmetric_{li}  +\momenta_A{}^i\momenta_B{}^j\tetrad^A{}_j\tetrad^B{}_i -\momenta_A{}^i\momenta_B{}^j\tetrad^A{}_i\tetrad^B{}_j \right]-\frac{\sqrt{\inducedmetric}}{2\kappa}{}^3 \mathbb{T}-\normalvector^A\partial_i \momenta_A{}^i \right] \\ &+\shift^j\left[ -\tetrad^A{}_j \partial_i \momenta_A{}^i-\momenta_A{}^i T^A{}_{ij} \right] -\lambda^{A}C_{A} -{}^{\mathcal{V}} \lambda_i\left[ -\dfrac{\kappa}{\sqrt{\inducedmetric}} \normalvector^A \momenta_A{}^i +T^B{}_{jk}\inducedmetric^{ik}\inducedmetric^{jl}\tetrad^A{}_l\eta_{AB} \right]\\
    &-{}^{\mathcal{A}} \lambda_{ij}\left[\dfrac{\kappa}{2\sqrt{\inducedmetric}}\tetrad^A{}_k(\momenta_A{}^j\inducedmetric^{ik}-\momenta_A{}^i\inducedmetric^{jk})  -\dfrac12 \inducedmetric^{ik}\inducedmetric^{jl}T^B{}_{kl}\normalvector_B \right].
    \label{eq.35}
\end{split}
\end{equation}
\twocolumngrid
In this expression, we have also imposed the Weitzenb\"{o}ck gauge, therefore all terms containing the spin connection $\omega^{A}{}_{Bi}$ vanish, and the type I Lorentz constraints proportional to $C_{AB}$ do not appear either. We are then left with a Hamiltonian that depends on the canonical variables $\theta^{A}{}_{i}$ and their conjugate momenta $\pi_{A}{}^{i}$. In the following sections, we obtain Hamilton's equations in terms of these variables.

\subsection{TEGR Hamilton's equations}
The Hamilton's equations for the spatial part of the tetrad are obtained as
\onecolumngrid
\begin{equation}
    \begin{split}
         \dot{\theta}^A{}_i &= \frac{\delta H}{\delta \pi_A{}^i}= \lapse\left( \frac{\kappa}{2\sqrt{\inducedmetric}}\left[ 2\momenta_B{}^j\tetrad^A{}_{[j}\tetrad^B{}_{i]} +\momenta_B{}^j \tetrad^A{}_k \tetrad^B{}_l\inducedmetric_{ij}\inducedmetric^{kl}\right]+\partial_{i}\normalvector^{A}  \right)\\
         & -\shift^j T^A{}_{ij} +\partial_i\left(\beta^{j}\tetrad^{A}_{j}\right)+{}^{\mathcal{V}}\lambda_i \frac{\kappa \normalvector^A}{\sqrt{\inducedmetric}} +{}^{\mathcal{A}}\lambda_{[ij]}\frac{\kappa \inducedmetric^{kj}\tetrad^A{}_k}{\sqrt{\inducedmetric}}  +\lambda_{[BC]}\eta^{AB}\tetrad^C{}_i.
         \label{HeqTetradFull}
    \end{split}
\end{equation}
In addition, the corresponding equations for the canonical momenta are
\begin{equation}
    \begin{split}
        -\dot{\pi}_A{}^i&= \dfrac{\delta H}{\delta \tetrad^{A}{}_{i} } = \dfrac{\lapse \kappa}{2\sqrt{\inducedmetric}}\left(  \momenta_A{}^j \momenta_B{}^i \tetrad^B{}_j-\momenta_A{}^i \momenta_B{}^j \tetrad^B{}_j+ \inducedmetric^{il}\inducedmetric_{jk}\momenta_A{}^j\momenta_B{}^k \tetrad^B{}_l\right)-\shift^i \partial_j \momenta_A{}^j + {}^{\mathcal{V}}\lambda^l T^B{}_{kl}(\tetrad_{A}{}^k\tetrad_{B}{}^{i}+\normalvector_{A}\normalvector_{B}\gamma^{ki} )\\
        &+\lapse \normalvector_A \inducedmetric^{ik}\tetrad^B{}_k\partial_j \momenta_B{}^j-{}^{\mathcal{V}}\lambda_j\frac{\kappa}{\sqrt{\inducedmetric}}\normalvector_A \inducedmetric^{ik}\tetrad^B{}_k \momenta_B{}^j -\dfrac12{}^{\mathcal{A}}\lambda^{lk}T^B{}_{kl}\normalvector_A \tetrad_B{}^i+\frac{\kappa}{\sqrt{\inducedmetric}}{}^{\mathcal{A}}\lambda_{[jk]}\momenta_A{}^j\inducedmetric^{ik} \\
        &+\frac{\kappa}{\sqrt{\inducedmetric}}{}^{\mathcal{A}}\lambda_{[lj]} \pi_B{}^j \tetrad^B{}_k (\tetrad_{A}{}^{l}\inducedmetric^{ik} + \tetrad_{A}{}^{k}\inducedmetric^{li} ) +\frac{\lapse \tetrad_A{}^i}{2}\left(-\frac{\sqrt{\inducedmetric}}{\kappa}{}^3 \mathbb{T}+\frac{\kappa}{2\sqrt{\inducedmetric}}\momenta_B{}^j\momenta_D{}^k\left(\tetrad^B{}_j\tetrad^D{}_k-\tetrad^B{}_k\tetrad^D{}_j-\inducedmetric_{jk}\inducedmetric^{ln}\tetrad^B{}_l\tetrad^D{}_n \right) \right)\\
        &-\frac{\kappa}{\sqrt{\inducedmetric}} {}^{\mathcal{V}}\lambda_j\tetrad_A{}^i\normalvector^B\momenta_B{}^j +{}^{\mathcal{V}}\lambda_j T^{B}{}_{kl}\tetrad_B{}^{k}(\tetrad_A{}^{j}\inducedmetric^{il} + \tetrad_{A}{}^{l} \inducedmetric^{ij} ) +\frac{\kappa}{\sqrt{\inducedmetric}} {}^{\mathcal{A}}\lambda_{[lj]} \inducedmetric^{kl}\tetrad_A{}^i\tetrad^B{}_k\momenta_B{}^j -\dfrac12{}^{\mathcal{A}}\lambda_{nj}T^B{}_{kl} \normalvector_B \left[\inducedmetric^{jl}(\tetrad_{A}{}^{n}\inducedmetric^{ik}\right.\\
        & \left.+ \tetrad_{A}{}^{k}\inducedmetric^{ni} ) + \inducedmetric^{nk}(\tetrad_{A}{}^{j} \inducedmetric^{il} + \tetrad_{A}{}^{l}\inducedmetric^{ji} ) \right] + \frac{\kappa \lapse}{2\sqrt{\inducedmetric}}\eta_{AC}\left(\inducedmetric^{kl}\tetrad^B{}_l\tetrad^D{}_k\tetrad^C{}_j \momenta_B{}^i\momenta_D{}^j -\inducedmetric_{jk}\inducedmetric^{lm}\inducedmetric^{in}\tetrad^B{}_n\tetrad^D{}_l\tetrad^C{}_m\momenta_B{}^j\momenta_D{}^k \right) \\
        &-2\partial_{l}\left(\shift^{[l}\momenta_A{}^{i]}-{}^{\mathcal{V}}\lambda^{[i} \tetrad_A{}^{l]}+\dfrac12 {}^{\mathcal{A}}\lambda^{[il]}\normalvector_A -\frac{\lapse\sqrt{\inducedmetric}}{\kappa}H_{CA}{}^{[mn][il]}T^C{}_{mn}\right) -\frac{\lapse \sqrt{\inducedmetric}}{\kappa}T^B{}_{kl}T^C{}_{mn} \left(\tetrad_{A}{}^{m}H\indices{^C^B_i_n_k_l}\right.\\
        &\left. + \tetrad_{A}{}^{n}H_{CB}{}^{mikl} + \tetrad_{A}{}^{k}H_{CB}{}^{mnil} + \tetrad_{A}{}^{l}H_{CB}{}^{mnki}  + \normalvector_C \normalvector_A \inducedmetric^{i[m}\inducedmetric^{n][k}\tetrad_{B}{}^{l]}  + \normalvector_B \normalvector_A \tetrad_{C}{}^{[m}\inducedmetric^{n][k}\inducedmetric^{l]i} \right).
        \label{HeqPiFull}
    \end{split}
\end{equation}
\twocolumngrid
These expressions do not consider that the primary constraints are zero on the constraint surface, as mentioned in the previous section. Once we impose the primary constraints in our computations, the final form of the evolution of tetrad and momenta components is considerably simplified.

Taking into account the previous discussion on spatial boundary terms, we acknowledge that there are several ways of writing these equations that will differ from each other through spatial derivatives of the spatial part of the tetrad, the lapse, and the shifts. However, none of them alters the dynamical structure or the physical content of the theory.
\subsection{On the equivalence between 3+1 GR and TEGR}
At this point, a natural question arises: why is it meaningful to study numerical relativity within the framework of TEGR, given that its equations of motion are classically equivalent to those of GR? If the dynamics are the same, one might expect the 3+1 decomposition to be identical as well. Now we aim to clarify these points.

There are multiple approaches in deriving the 3+1 equations of motion in TEGR. One method involves starting from the TEGR Lagrangian, constructed from the torsion scalar $\mathbb T$, performing the 3+1 decomposition at the Lagrangian level, and varying with respect to the spatial tetrad $\theta^{A}{}_{i}$. Alternatively, one can apply the decomposition directly to the field equations that are equivalent to Einstein's equations. It should yield the same result as GR, albeit expressed in the language of tetrads. 

Nonetheless, an important subtlety arises: the 3+1 decomposition of the torsion scalar $\mathbb T$ is not unique. Specifically, terms involving spatial derivatives can be integrated by parts in different ways, leading to ambiguities in the resulting 3+1 Lagrangian and, consequently, in the equations of motion. This discussion appears in~\cite{Guzman:2023oyl} for the symmetric teleparallel equivalent of GR, and previous work for TEGR \cite{Pati:2022nwi,Blixt:2020ekl}. In these works, there are hints that such ambiguities can affect the non-dynamical structure of the equations. In this case, alternative approaches to GR do not introduce significant observational advantages at the classical level, but offer an exploration of reformulations of general relativity that could influence the setup for numerical relativity. This stands in contrast to critiques, such as in \cite{Golovnev:2024lku}, which argues the lack of influence of the geometric entities in the physical equations of motion. On the contrary, the flexibility in reformulating GR through teleparallel geometry may open new avenues for computational implementation with unexplored consequences.

\section{Linearisation of the evolution equations of TEGR}
\label{sec:Lin}

To date, the studies addressing the principal symbol analysis and properties of the evolution equations in TEGR are scarce \cite{Peshkov:2022cbi}. In this section, we aim to contribute filling in this gap by presenting a simple yet illustrative example that explores the linearised evolution equations and prepares them for the analysis of the principal symbol.

Our approach follows the methodology outlined by Alcubierre in \cite{article}, where the linearised ADM equations of GR are analysed. We adopt this procedure to the TEGR framework, considering it as a toy model that lays the groundwork for more comprehensive investigations of hyperbolicity in the general case. Despite its simplicity, this example captures several subtleties and complications associated with the choice of tetrad orientation in a generic gauge, an aspect that can either facilitate or complicate the analysis. We also demonstrate how the constraints can be used to improve the system of differential equations.
\subsection{Linearising the 3+1 decomposition of the tetrad}
In \cite{article}, linear perturbations in the ADM equations in general relativity around a flat spacetime with vanishing shift vector, i.e. $\beta_{i} = 0$, were studied. The non-vanishing components and the linearity assumption implies that the lapse and induced metric can be decomposed as
\begin{equation}
\alpha = 1+a, \qquad \gamma_{ij} = \delta_{ij} + h_{ij},
\label{GRconditions}
\end{equation}
with $a,h_{ij}\ll1$. In order to reproduce this expansion in the tetrad formalism, we assume a perturbative expansion of the tetrad components as
\begin{equation}
\tetrad^{A}{}_{\mu} = \delta^{A}_{\mu} +  h^{A}{}_{\mu} 
\label{LinTetrad}
\end{equation}
with $h^{A}{}_{\mu} \ll 1$. In order to match the metric components in Eq.~\eqref{GRconditions} with this expansion, we  extract the temporal and spatial components of the linearised tetrad in Eq.~\eqref{LinTetrad} in the most general case, for both spacetime and flat indices. This gives us
\begin{align}
    \begin{split}
& \theta^{0}{}_{0} = 1 + h^{0}{}_{0}, \qquad \  \theta^{0}{}_{i} = h^{0}{}_{i}, \\
& \theta^{a}{}_{0} = h^{a}{}_{0}, \qquad \qquad \theta^{a}{}_{i} = \delta^{a}_{i} + h^{a}{}_{i}.
\end{split}
\end{align}
Notice that the previous condition $g_{0i} = \beta_i = 0$ implies $\theta^{A}{}_0 = \alpha \xi^{A}$, and at the level of perturbations, that
\begin{equation}
\begin{split}
\delta g_{0i} & = \eta_{AB} h^{A}{}_{0} h^{B}{}_{i} = \eta_{00} h^{0}{}_{0} h^{0}{}_{i} + \delta_{kj} h^{k}{}_{0} h^{j}{}_{i}\\
& = - h^{0}{}_{0} h^{0}{}_{i} + h^{j}{}_{0} h^{j}{}_{i}.
\end{split}
\end{equation}
Finally, it is also true that $g^{0i} = \beta^{i}/\alpha^2 = 0$.

Additionally, in order to match the lapse perturbations for the metric, we observe that $\delta g_{00} = -2a$, since
$$g_{00} = - \alpha^2 = -(1+a)^2 \approx -(1+2a).$$
On the other hand,
$$\delta g_{00} = 2 \eta_{A0} h^{A}{}_{0} = 2 \eta_{00} h^{0}{}_{0} = -2h^{0}{}_{0},$$
therefore, it is satisfied that $a = h^{0}{}_{0}$. Furthermore, we can write the inverse tetrad components and their expansion as 
\begin{equation}
e_{A}{}^{\mu} = \delta_{A}^{\mu} + h_{A}{}^{\mu},
\end{equation}
with $h_{A}{}^{\mu} \ll 1$. For completeness, we find the relations among the perturbations of the components of the inverse tetrad $h\indices{_A^\mu}$ and the perturbations of the components of the tetrad $h\indices{^A_\mu}$. Since $e_{A}{}^{\mu} = g^{\mu\nu} \eta_{AB} \theta^{B}{}_{\nu}$, we expand both spacetime and flat indices in temporal and spatial components, obtaining the following results 
\begin{equation}
e_{i}{}^{j} = \eta_{ik} (g^{j0} \theta^{k}{}_{0} + g^{jl} \theta^{k}{}_{l}) = \delta^{j}_{i} +  h_{i}{}^{j} + \dfrac{1}{\alpha^2} \eta_{ik} \beta^{j} h^{k}{}_{0},
\end{equation}
\begin{equation}
    e_{0}{}^{i}  = \eta_{00}( g^{i0} \theta^{0}{}_{0} + g^{ij} \theta^{0}{}_{j}) = -\dfrac{\beta^{i}}{\alpha^2}\left(1 + h^{0}{}_{0} \right) - \gamma^{ij} h^{0}{}_{j} ,
\end{equation}
\begin{equation}
    e_{0}{}^{0} = \eta_{00}( \theta^{0}{}_{0} g^{00} +  \theta^{0}{}_{i} g^{0i}) = \dfrac{1}{\alpha^2}(1+h^{0}{}_{0})-\dfrac{1}{\alpha^2} \beta^{i} h^{0}{}_{i}.
\end{equation}
It is also useful to write explicitly the linearisation of the components of the torsion tensor $T\indices{^A_\mu_\nu}$,
\begin{equation}
    T\indices{^0_0_i} = \partial_{0} \tetrad \indices{^0_i} - \partial_{i} \tetrad \indices{^0_0} \approx  \partial_{0} h \indices{^0_i} - \partial_{i} h \indices{^0_0},
\end{equation}
\begin{equation}
   T\indices{^0_i_j}  = \partial_i \tetrad^{0}{}_{j} - \partial_{j} \tetrad^{0}{}_{i} \approx \partial_i h^{0}{}_{j} - \partial_{j} h^{0}{}_{i},
\end{equation}
\begin{equation}
    T\indices{^a_0_i} = \partial_{0} \tetrad \indices{^a_i} - \partial_{i} \tetrad \indices{^a_0} \approx \partial_{0} h \indices{^a_i} - \partial_{i} h \indices{^a_0},
\end{equation}
\begin{equation}
    T\indices{^a_i_j} = \partial_i \tetrad^{a}{}_{j} - \partial_{j} \tetrad^{a}{}_{i} \approx \partial_i h^{a}{}_{j} - \partial_j h^{a}{}_{i},
    \label{lintorsion}
\end{equation}
while the remaining components are zero. We remark that the torsion tensor is linear in tetrad perturbations. Consequently, any occurrence of the torsion tensor in quadratic combinations in the evolution equations can be considered as a second-order contribution in the perturbative expansion and will be neglected in this treatment. 

Last but not least, we compute the linearisation of the normal vector 
\begin{equation}
      \xi^{A}  = -\dfrac16 \epsilon\indices{^A_B_C_D} \tetrad^{B}{}_i \tetrad^{C}{}_j \tetrad^{D}{}_k \epsilon^{ijk}.
\end{equation}
By construction, the only non-zero component of $\xi^{A}$ is the temporal component $A=0$, which allows us to write $\epsilon\indices{^A_B_C_D} =  E^{A} \epsilon_{bcd}$, where $E^{A}$ is a unit normal vector. With this, we expand the normal vector as
\onecolumngrid
\begin{equation}
\begin{split}
    \xi^{A} & = -\dfrac16 \epsilon\indices{^A_B_C_D} \tetrad^{B}{}_i \tetrad^{C}{}_j \tetrad^{D}{}_k \epsilon^{ijk} \approx -\dfrac16 E^{A} \epsilon_{bcd} \tetrad^{b}{}_i \tetrad^{c}{}_j \tetrad^{d}{}_k \epsilon^{ijk}, \\
    &  \approx -\dfrac16 E^{A} \epsilon_{bcd} (\delta^{b}_i + h^{b}{}_i) ( \delta^{c}_j + h^{c}{}_{j} ) ( \delta^{d}_k + h^{d}{}_{k} ) \epsilon^{ijk}, \\
    & \approx -\dfrac16 E^{A} \epsilon_{bcd} \epsilon^{ijk} \delta^{b}_i \delta^{c}_j \delta^{d}_k -\dfrac16 E^{A} \epsilon_{bcd} \epsilon^{ijk} (h^{b}{}_{i} \delta^{c}_j \delta^{d}_k + h^{c}{}_{j}\delta^{b}_i\delta^{d}_k +  h^{d}{}_{k}\delta^{b}_i \delta^{c}_j  ), \\
     & \approx -\dfrac16 E^{A} \epsilon_{bcd} \epsilon^{bcd}   - \dfrac12 E^{A} \epsilon_{bcd} \epsilon^{icd} h^{b}{}_{i}.
\end{split}
\end{equation}
\twocolumngrid
Using $\epsilon_{ijk}\epsilon^{ijn}=2\delta^{n}_{k}$ and $\epsilon_{bcd}\epsilon^{bcd}=6$, we get
\begin{equation}
    \xi^A\approx - E^A-E^A \sum_{b=1}^{3} h^{b}{}_{b} =- E^A[ 1+ tr(h) ].
    \label{temporal}
\end{equation}
Our next step is to use all the previous relations for linearising Hamilton's equations of TEGR.
\subsection{Linearisation of Hamilton's equations in TEGR}
We begin by linearising the evolution equations for the tetrad components in Eq.~\eqref{HeqTetradFull}. We start by setting all Lagrange multipliers to zero, which means that we neglect the contribution of constraints to the evolution equations. However, we keep in mind that they can later be reintroduced to alter the properties of the principal symbol of the system. The resulting set of equations are 
\onecolumngrid
\begin{equation}\label{tetraddotsimpl1}
\dot{\theta}^A{}_i=\frac{\delta H}{\delta \pi_A{}^i} = \lapse\left( \frac{\kappa}{2\sqrt{\inducedmetric}}\left[ 2\momenta_B{}^j\tetrad^A{}_{[j}\tetrad^B{}_{i]} +\momenta_B{}^j \tetrad^A{}_k \tetrad^B{}_l\inducedmetric_{ij}\inducedmetric^{kl}\right]+\partial_{i}\normalvector^{A} \right).
\end{equation}
\twocolumngrid
A subtle distinction from the evolution equation for the intrinsic metric in GR is that the evolution equations in TEGR for the tetrad include an extra term $\partial_i \xi^{A}$. This feature has the potential to influence the principal symbol analysis that is not present in GR.

It is important to emphasise that, from now on, we will treat the conjugate momenta $\pi_{A}^{i}$ as a first-order term in perturbations. This assumption stems from the analogy with the analysis in general relativity \cite{article}, where the conjugate momenta of the metric $\pi_{ij}$ are proportional to the extrinsic curvature $K_{ij}$, which is a first order term as it is proportional to the time derivative of the spatial metric $\gamma_{ij}$. Since $\gamma_{ij} = \eta_{ij} + h_{ij}$, it is implied that $K_{ij} \approx \partial_0 h_{ij}$ is linear in perturbations.

With the first two terms in Eq.~\eqref{tetraddotsimpl1} first-order, all other terms different from the momenta $\pi_{A}{}^{i}$ must be evaluated at the background level. More specifically, if we consider the evolution equation for $\dot{\theta}^{0}{}_{i}$, we find that the perturbations $h^{0}{}_{i}$ evolve as
\begin{equation}
\begin{split}
\dot{h}^0{}_i & = \frac{1}{2}\left[ 2\momenta_B{}^j\delta^0{}_{[j}\delta^B{}_{i]} +\momenta_B{}^j \delta^0{}_k \delta^B{}_l\delta_{ij}\delta^{kl}\right]  - \partial_{i}h^{j}{}_{j} \\ 
& = -  \partial_{i}h^{j}{}_{j},
\end{split}
\end{equation}
where from now on we set $\kappa=1$ and $\sqrt{\gamma} = 1$. From here we see that all the terms vanish, except for those that correspond to the spatial derivatives of the trace of the tetrad perturbation. In Cartesian coordinates, the equations reduce to
\begin{equation}
\dot{h}^{0}{}_{x} + \partial_x( h^{x}{}_{x} + h^{y}{}_{y} + h^{z}{}_{z} ) = 0,
\end{equation}
while $\dot{h}^{0}{}_{y} = \dot{h}^{0}{}_{z} = 0$.

Meanwhile, the evolution equations for the spatial ($A=a$) components of the tetrad perturbations $h^{a}{}_{i}$ take the form
\begin{equation}
\dot{h}^{a}{}_{i} = \dfrac{1}{2} \left[ \pi_{b}{}^{j}(\delta^{a}_{j} \delta^{b}_{i} - \delta^{a}_{i} \delta^{b}_{j} ) + \pi_{b}{}^{j} \delta^{a}_{k} \delta^{b}_{l} \delta_{i j} \delta^{k l} \right].
\label{hdot_full}
\end{equation}
These equations are linear in the momenta and contain no spatial derivatives of any variables. They are analogous to the evolution equations for the intrinsic metric in GR, since both sets of equations depend linearly on the conjugate momenta $\pi_{ij}$ and $\pi^{i}{}_{A}$, respectively. These equations do not include spatial derivatives, therefore they are not directly useful for studying properties of the principal symbol. Nevertheless, we will need them to complete the system of differential equations, and consequently we write the resulting linearised equations in components here. Let us consider all the possible spatial dimensions for the indices $a$ and $i$ in Cartesian coordinates in Eq.~\eqref{hdot_full}. For the diagonal components $a = i$, we obtain
\begin{equation}
\begin{split}
\partial_t h^{x}{}_{x} - \dfrac12 \pi_{x}{}^{x} + \dfrac12 \pi_{y}{}^{y} + \dfrac12 \pi_{z}{}^{z} & =  0, \\
\partial_t h^{y}{}_{y} + \dfrac12 \pi_{x}{}^{x} -  \dfrac12 \pi_{y}{}^{y} + \dfrac12 \pi_{z}{}^{z} & = 0, \\
\partial_t h^{z}{}_{z} + \dfrac12 \pi_{x}{}^{x} + \dfrac12 \pi_{y}{}^{y} - \dfrac12 \pi_{z}{}^{z}  & = 0,
\end{split}
\label{dhdt_diag}
\end{equation}
while for the off-diagonal components
\begin{equation}
\begin{split}
& \partial_t h^{x}{}_{y} - \dfrac12 \pi_{x}{}^{y} - \dfrac12 \pi_{y}{}^{x} = 0, \quad \partial_t h^{x}{}_{z} - \dfrac12 \pi_{x}{}^{z} - \dfrac12 \pi_{z}{}^{x} = 0, \\
& \partial_t h^{y}{}_{x} - \dfrac12 \pi_{y}{}^{x} - \dfrac12 \pi_{x}{}^{y} = 0, \quad \partial_t h^{y}{}_{z} - \dfrac12 \pi_{y}{}^{z} - \dfrac12 \pi_{z}{}^{y} = 0, \\
& \partial_t h^{z}{}_{x} - \dfrac12 \pi_{z}{}^{x} - \dfrac12 \pi_{x}{}^{z} = 0, \quad \partial_t h^{z}{}_{y} - \dfrac12 \pi_{z}{}^{y} - \dfrac12 \pi_{y}{}^{z} = 0.
\end{split}
\label{dhdt_offdiag}
\end{equation}
Similarly, for the tetrad components, we linearise the evolution equations for the conjugate momenta \eqref{HeqPiFull}. We omit all Lagrange multipliers, obtaining
\onecolumngrid
\begin{equation}
    \begin{split}
        -\dot{\pi}_A{}^i &= \frac{\lapse\kappa}{2\sqrt{\inducedmetric}}\left(  \momenta_A{}^j \momenta_B{}^i \tetrad^B{}_j-\momenta_A{}^i \momenta_B{}^j \tetrad^B{}_j+ \inducedmetric^{il}\inducedmetric_{jk}\momenta_A{}^j\momenta_B{}^k \tetrad^B{}_l\right) +\lapse \normalvector_A \inducedmetric^{ik}\tetrad^B{}_k\left(\partial_j \momenta_B{}^j\right)+2\partial_{l}\left(\frac{\lapse\sqrt{\inducedmetric}}{\kappa}H_{CA}{}^{[mn][il]}T^C{}_{mn}\right) \\
       & +\frac{\kappa\lapse }{4\sqrt{\inducedmetric} }\left( \tetrad_A{}^i\momenta_B{}^j\momenta_D{}^k\left(\tetrad^B{}_j\tetrad^D{}_k-\tetrad^B{}_k \tetrad^D{}_j-\inducedmetric_{jk}\tetrad^{Bn}\tetrad^D{}_n \right) \right) + \frac{\kappa \lapse}{2\sqrt{\inducedmetric}}\eta_{AC}\left(\tetrad^{Bl}\tetrad^D{}_k\tetrad^C{}_j \momenta_B{}^i\momenta_D{}^j-\tetrad^{Bi}\tetrad^{Dm}\tetrad^C{}_m\momenta_{Bk}\momenta_D{}^k \right)\\& -\frac{\lapse }{2\kappa  }\left(\tetrad_A{}^i \sqrt{\inducedmetric}{}^3 \mathbb{T} \right)-\frac{\lapse \sqrt{\inducedmetric}}{\kappa}T^B{}_{kl}T^C{}_{mn}\left( \tetrad_{A}{}^{m}H_{CB}{}^{inkl} + \tetrad_{A}{}^{n}H_{CB}{}^{mikl}  + \tetrad_{A}{}^{k}H_{CB}{}^{mnil} + \tetrad_{A}{}^{l}H_{CB}{}^{mnki} \right.\\
        &\left. + \normalvector_C \normalvector_A \inducedmetric^{i[m}\inducedmetric^{n][k}\tetrad_{B}{}^{l]}  + \normalvector_B \normalvector_A \tetrad_{C}{}^{[m}\inducedmetric^{n][k}\inducedmetric^{l]i} \right),
        \label{pidotsimpl1}
    \end{split}
 \end{equation}
\twocolumngrid
where $^3\mathbb{T}$ is the spatial part of the torsion scalar
\begin{equation}
\begin{split}
    {}^3\mathbb{T}&=H_{AB}{}^{ijkl}T^A{}_{ij}T^B{}_{kl},
    \label{3TorSpatial}
\end{split}
\end{equation}
with
\begin{equation}
    H_{AB}{}^{ijkl}= -\frac{1}{4}\eta_{AB} \inducedmetric^{k[i}\inducedmetric^{j]l}+\frac{1}{2}\tetrad_B{}^{[i}\inducedmetric^{j][k}\tetrad_A{}^{l]}-\tetrad_A{}^{[i}\inducedmetric^{j] [k}\tetrad_B{}^{l]}.
\end{equation}
As shown in Eq. \eqref{lintorsion}, the torsion tensor depends linearly on the tetrad perturbations. Consequently, the spatial torsion scalar ${}^3\mathbb{T}$, being quadratic in the torsion tensor, is of second order in perturbations and can therefore be neglected at linear order. The same argument applies to the terms in Eq. \eqref{pidotsimpl1} that are proportional to $T^{B}{}_{kl} T^{C}{}_{mn}$. What remains is the following
\begin{align}
    \begin{split}
        -\dot{\pi}_A{}^i &= \dfrac{\kappa\lapse}{2\sqrt{\inducedmetric}}\left(  \inducedmetric^{il}\inducedmetric_{jk}\momenta_A{}^j\momenta_B{}^k \tetrad^B{}_l+\momenta_A{}^j \momenta_B{}^i \tetrad^B{}_j\right.\\ 
        &\left.-\momenta_A{}^i \momenta_B{}^j \tetrad^B{}_j\right)+ \frac{\kappa \lapse}{2\sqrt{\inducedmetric}}\eta_{AC}\left(\tetrad^{Bl}\tetrad^D{}_k\tetrad^C{}_j \momenta_B{}^i\momenta_D{}^j\right.\\
        &-\left.\tetrad^{Bi}\tetrad^{Dm}\tetrad^C{}_m\momenta_{Bk}\momenta_D{}^k \right)+\lapse \normalvector_A \inducedmetric^{ik}\tetrad^B{}_k\left(\partial_j \momenta_B{}^j\right)\\ 
        &+\frac{\kappa\lapse }{4\sqrt{\inducedmetric} }\left( \tetrad_A{}^i\momenta_B{}^j\momenta_D{}^k\left(\tetrad^B{}_j\tetrad^D{}_k-\tetrad^B{}_k \tetrad^D{}_j\right.\right.\\
        &\left.\left.-\inducedmetric_{jk}\tetrad^{Bn}\tetrad^D{}_n \right) \right)+\dfrac{2}{\kappa}\partial_{l}\left( \lapse\sqrt{\inducedmetric}H_{CA}{}^{[mn][il]}T^C{}_{mn}\right).\\
        \label{pidotsimpl2}
    \end{split}
 \end{align}
The expressions multiplying the spatial derivatives of the conjugate variables $\pi_{A}{}^{i}$ and $\theta^{A}{}_{i}$ are the coefficients that constitute the principal symbol, and influence their hyperbolicity properties. In analogy with GR, we also discard terms that are quadratic in $\pi_{A}{}^{i}$, since they contribute at second order in perturbations. These simplifications yield a significantly more tractable expression
\begin{equation}
        -\dot{\pi}_A{}^i = \lapse \normalvector_A \inducedmetric^{ik}\tetrad^B{}_k\partial_j \momenta_B{}^j+\dfrac{2}{\kappa}\partial_{l}\left( \lapse\sqrt{\inducedmetric}H_{CA}{}^{[mn][il]}T^C{}_{mn}\right).
        \label{pidotsimpl4}
 \end{equation}
We are interested in obtaining a simpler form for these equations in the cases $A=0$ and $A=a$. The component $A=0$ of Eq.~\eqref{pidotsimpl4} gives the following contribution
\begin{equation}
        -\dot{\pi}_0{}^i = \lapse \normalvector_0 \inducedmetric^{ik}\tetrad^B{}_k\partial_j \momenta_B{}^j +\dfrac{2}{\kappa}\partial_{l}\left( \lapse\sqrt{\inducedmetric} H_{C0}{}^{[mn][il]}T^C{}_{mn}\right).\\
        \label{pidotsimpl0}
 \end{equation}
We will demonstrate intermediate computations that will simplify this equation. The relevant components of the tensor $H$ to be linearised are
\begin{equation}
H\indices{_C_0^[^m^n^]^[^i^l^]}=\frac{1}{2}\tetrad_0{}^{[m}\inducedmetric^{n][i}\tetrad_C{}^{l]}-\frac{1}{4}\eta_{C0} \inducedmetric^{i[m}\inducedmetric^{n]l}-\tetrad_C{}^{[m}\inducedmetric^{n] [i}\tetrad_0{}^{l]}. 
 \end{equation}
We extract the temporal components of this tensor for $C=0$, and linearise to get
 \begin{equation}
 H_{00}{}^{[mn][il]} = \dfrac14 \delta^{i[m}\delta^{n]l} - \dfrac12 h^{0}{}_{p} h^{0}{}_{q} \delta^{p[m} \delta^{n][i} \delta^{l]q},
 \end{equation}
while for the spatial part $C=k$, we obtain
   \begin{equation}
H_{k0}{}^{[mn][il]} \approx h^{0}{}_{p} \left(-\frac{1}{2}\delta^{p[m}\delta^{n][i}\delta_k{}^{l]} + \delta^{p[l} \delta^{i][n} {\delta^{m]}}_{k} \right).
\label{Hk0_1storder}
 \end{equation}
We see that all these components of the tensor $H$ are linear or quadratic in tetrad perturbations $h^{0}{}_{i}$, except for the first term in $H_{00}{}^{[mn][il]}$. Since the components of $H_{C0}{}^{[mn][il]}$ are multiplied by the torsion tensor, the only background contributions that plays a role is the term $H_{00}{}^{[mn][il]}$. 
Therefore, we write
\begin{equation}
- \dot{\pi}_0{}^{i} = \xi_0 \delta^{ik} \partial_j \pi_{k}{}^{j} + \dfrac14 ( \delta^{im} \delta^{nl} - \delta^{in} \delta^{ml} )\partial_l (\partial_m h^0{}_n - \partial_n h^0{}_m ).
\end{equation}
Now, for deriving the linearised evolution equation for $\pi_{a}{}^{i}$ in Eq.~\eqref{pidotsimpl4} (with only spatial $a$), we notice that the only terms that contribute at the first order are
 \begin{equation}
 \begin{split}
        -\dot{\pi}_a{}^i & = \dfrac{2}{\kappa} H_{Ca}{}^{[mn][il]} \partial_{l}T^C{}_{mn} \\
        & = \dfrac{2}{\kappa} H_{0a}{}^{[mn][il]} \partial_{l}T^0{}_{mn} + \dfrac{2}{\kappa} H_{ca}{}^{[mn][il]} \partial_{l}T^c{}_{mn}.
\end{split}
\label{pidotsimpl5}
 \end{equation}
Nonetheless, we have seen in Eq. \eqref{Hk0_1storder} that the term $H_{0c}{}^{[mn][il]}$ contributes purely at first order and when multiplied by the torsion tensor, the product becomes second order. Therefore, the only term that survives in Eq.~\eqref{pidotsimpl5} is $H_{ca}{}^{[mn][il]}$. We linearise this expression, obtaining
 \begin{equation}
 \begin{split}
H_{ca}{}^{[mn][il]} & =  \dfrac12 \theta_{a}{}^{[m} \gamma^{n][i}\theta_{c}{}^{l]} - \theta_{c}{}^{[m} \gamma^{n][i} \theta_{a}{}^{l]} -\dfrac14 \eta_{ca} \gamma^{i[m}\gamma^{n]l}\\
& \approx  \dfrac12 \delta^{[m}_{a} \delta^{n][i}\delta_c^{l]} - \delta_c^{[m}\delta^{n][i}\delta_a^{l]} -\dfrac14 \eta_{ca} \delta^{i[m}\delta^{n]l}.
\end{split}
 \end{equation}
Henceforth we conclude that the only contribution to the time derivative of ${\pi_a}^i$ can be written as \footnote{We have used that $ A^{[mn]}(\partial_m h^{C}{}_n - \partial_n h^{C}{}_{m} ) = 2 A^{[mn]} \partial_m h^{C}{}_{n}$, and $\kappa = 1$.}
\begin{equation}
\begin{split}
-\dot{\pi}_a{}^i  = &  ( 2 \delta^{[m}_{a} \delta^{n][i}\delta_c^{l]}- \eta_{ca} \delta^{i[m}\delta^{n]l} \\
& - 4\delta_c^{[m}\delta^{n][i}\delta_a^{l]}  ) \partial_l \partial_m h^{c}{}_n.
\end{split}
\label{pidotspatial}
\end{equation}
Our next step is to expand the antisymmetrised terms and specialise to Cartesian coordinates,  and consider that all perturbations depend only on $x$, therefore the partial derivatives with respect to the remaining coordinates vanish. We obtain the following set of differential equations
\begin{equation}
\begin{split}
 -\dot{\pi}_{y}{}^{y} & =   -\partial_x \partial_{x}h^{z}{}_{z}, \\ 
 -\dot{\pi}_{y}{}^{z} & = \dfrac{1}{2} \partial_x \partial_x h^{y}{}_{z} + \dfrac{1}{2} \partial_x \partial_x h^{z}{}_{y}, \\
 -\dot{\pi}_{z}{}^{y} & = \dfrac{1}{2} \partial_x \partial_x h^{z}{}_{y} + \dfrac{1}{2} \partial_x \partial_x h^{y}{}_{z}, \\
 -\dot{\pi}_{z}{}^{z} & = -  \partial_x \partial_{x} h^{y}{}_{y},
\label{pidoteqs}
\end{split}
\end{equation}
while the remaining equations for the remaining combination of coordinates are zero, i.e. $\dot{\pi}_{b}{}^{j} = 0$. 
\subsection{Linearised TEGR constraints}
For a complete analysis, we consider also the linearised version of the TEGR constraints. These equations will be helpful later to modify the structure of the differential equations and therefore their properties. We start with the six Lorentz constraints in Eq.~\eqref{LorC1} and Eq.~\eqref{LorC2}, which are generators of infinitesimal boosts and rotations in the tetrad, respectively \cite{Ferraro:2016wht}. The linearised version of the set of constraints in Eq.~\eqref{LorC1} is
\begin{equation}
    -\pi_{0}{}^{i} + \eta_{ab} \delta^{b}_{l} \delta^{jl} \delta^{ik}( \partial_j h^{a}{}_{k} - \partial_k h^{a}{}_{j} ) = 0. 
\end{equation}
We write the three possible values of the index $i$, obtaining
\begin{equation}
\begin{split}
C^{x} & = \pi_{0}{}^{x}  - \partial_x h^{y}{}_{y} - \partial_x h^{z}{}_{z} = 0, \\
C^{y} & = \pi_{0}{}^{y} + \partial_x h^{x}{}_{y} = 0,\\
C^{z} & = \pi_{0}{}^{z} + \partial_x h^{x}{}_{z} = 0.
\end{split}
\end{equation}
For the set of constraints in Eq.~\eqref{LorC2}, the linearisation gives the general form
\begin{equation}
C^{ij} = -\dfrac14 \delta^a_k ( \pi_k{}^i \gamma^{jk} - \pi_k{}^j \gamma^{ik} ) - \dfrac12 \delta^{ik} \delta^{jl} T^{0}{}_{kl} = 0, 
\label{LinLorentz2}
\end{equation}
and after evaluating in Cartesian coordinates, it follows the result (after normalisation by a factor of $1/2$)
\begin{equation}
\begin{split}
    C^{xy} & = - C^{yx} = \dfrac{1}{2}( \pi_{y}{}^{x} - \pi_{x}{}^{y} ) + \partial_x h^{0}_y = 0, \\
    C^{xz} & = - C^{zx} = \dfrac{1}{2}( \pi_{z}{}^{x} - \pi_{x}{}^{z} ) + \partial_x h^0{}_{z}  = 0, \\
    C^{yz} & = - C^{zy} = \dfrac{1}{2}( \pi_{z}{}^{y} - \pi_{y}{}^{z} )  = 0.
\end{split}
\label{Lor2lin}
\end{equation}
In the $C^{yz}$ there are no spatial derivatives of $h^{0}{}_{i}$ since the perturbations do not depend on $y$ and $z$, therefore their derivatives $\partial_y, \partial_z$ vanish. Since these constraints are antisymmetric in their two indices, there are only three linearly independent constraints. We also consider the momenta constraints in TEGR in the Weitzenb\"{o}ck gauge \cite{Pati:2022nwi}
\begin{equation}
    -\theta^{A}{}_{i} \partial_{j} \pi_{A}{}^{j} - \pi_{A}{}^{j} T^{A}{}_{ji} = 0,
\end{equation}
and notice that after linearisation only the first term survives, therefore the momenta constraints become
\begin{equation}
-\theta^{a}{}_{i} \partial_j \pi_{a}{}^{j} = 0.
\end{equation}
These equations correspond to three conditions on the spatial derivatives of the components of the canonical momenta. After adopting Cartesian coordinates and taking coordinate dependence only along $x$ direction, we get
\begin{equation}
\partial_x \pi_{x}{}^{x} = 0, \qquad \partial_x \pi_{y}{}^{x} = 0, \qquad \partial_x \pi_{z}{}^{x} = 0.
\end{equation}
Finally, we consider the linearised version of the Hamiltonian constraint, which was also derived in \cite{Pati:2022nwi} for TEGR, and can be written as
\begin{equation}
\begin{split}
     \mathcal{H} &= \dfrac{\kappa}{2\sqrt{\gamma}} \left( \dfrac12 \theta^{A}{}_{i} \theta^{B}{}_{j}(\pi_{B}{}^{i} \pi_{A}{}^{j} - \pi_{A}{}^{i} \pi_{B}{}^{j} ) \right. \\ 
  &\left.+ \dfrac12 \theta^{A}{}_{k} \theta^{B}{}_{j} \pi_{A}{}^{i} \pi_{B}{}^{l} \gamma^{jk} \gamma_{li} \right) - \dfrac{\sqrt{\gamma}}{2\kappa} {}^{(3)} \mathbb T - \xi^{A} \partial_i \pi_{A}{}^{i}.
\end{split}
\end{equation}
The last term is the only contribution at first order, which gives, in Cartesian coordinates
\begin{equation}
    \mathcal{H}_0 = \partial_i \pi_0{}^{i} = \partial_x \pi_0{}^{x} \approx 0.
\end{equation}

With this, we start our exploration of the mathematical properties of the principal symbol of $3+1$ TEGR equations, and identify if the system is well-posed under our assumptions.
\section{Hyperbolicity of TEGR dynamical equations}
\label{sec:Hyperb}
\subsection{Definition of (strong) hyperbolicity}
Consider a one-dimensional first-order system of equations for the set of variables $u_i$
\begin{equation}
    \partial_t u_i + \sum_{j} M_{ij} \partial_x u_j = q_i(u),
    \label{diffeqgeneral}
\end{equation}
for $i \in \{ 1, \ldots, N_u \}$, $N_u$ the total number of fields. For simplicity we consider $q_i = 0$, but in principle, it could also be an arbitrary function of $u_i$. $M_{ij}$ is the principal symbol or Jacobian matrix, and we will be interested in its eigenvalues $\lambda_i$ and eigenvectors $v_i$. The system of equations is \textit{hyperbolic} if all the eigenvalues $\lambda_i$ are real. The system is \textit{strongly hyperbolic} if the all the eigenvalues are real, and the principal symbol $M_{ij}$ admits a complete set of eigenvectors. If the eigenvalues are real but there is not a full set of eigenvectors, the system is called \textit{weakly hyperbolic}.
\subsection{First-order reduction}
We first need to reduce the linearised TEGR evolution system in Eq.~\eqref{pidoteqs} into first-order in space and time. 

In the analysis for GR \cite{article}, we have
\begin{equation}
    \partial_t a = -\text{tr}K = -K.
\end{equation}
In order to have a strictly first-order system, auxiliary variables are introduced,
\begin{equation}
    A_i = \partial_i a, \qquad D_{ijk} = \dfrac12 \partial_i h_{jk}.
\end{equation}
On the contrary, the linearised TEGR equations do not include a differential equation for the lapse or its spatial derivative. We also note that in TEGR, we have second order spatial derivatives of the tetrad instead of the metric in GR. Therefore, we introduce auxiliary variables for the spatial derivatives of the tetrad perturbations. For a 1-dimensional analysis, we restrict to dependence on the Cartesian coordinate $x$ to get
\begin{equation}
   D^{a}{}_{i} := \partial_x h^{a}{}_{i}, \qquad D^{0}{}_{i} = \partial_x h^{0}{}_{i}.
    \label{redef}
\end{equation}
Taking the $x$-derivative of the system of equations in Eq.~\eqref{dhdt_diag} and Eq.~\eqref{dhdt_offdiag}, and rewriting it in terms of $D^a{}_{i}$, yields the following first-order system: 
\begin{equation}\label{thetadoteqs_rew}
    \begin{split}
     \dot D^{x}{}_{x} - \dfrac12 \partial_x \pi_{x}{}^{x} + \dfrac12\partial_x \pi_{y}{}^{y}  + \dfrac12\partial_x \pi_{z}{}^{z} &  = 0,\\
     \dot D^{x}{}_{y} - \dfrac12 \partial_x\pi_{x}{}^{y} - \dfrac12 \partial_x \pi_{y}{}^{x} & = 0,\\
    \dot D^{x}{}_{z} - \dfrac12 \partial_x\pi_{x}{}^{z} - \dfrac12\partial_x \pi_{z}{}^{x} & = 0, \\
    \dot D^{y}{}_{x} - \dfrac12 \partial_x\pi_{y}{}^{x} - \dfrac12\partial_x \pi_{x}{}^{y} &= 0, \\
   \dot D^{y}{}_{y}   + \dfrac12 \partial_x \pi_{x}{}^{x}  - \dfrac12\partial_x \pi_{y}{}^{y} +  \dfrac12\partial_x \pi_{z}{}^{z} & = 0, \\
\dot D^{y}{}_{z} - \dfrac12 \partial_x\pi_{y}{}^{z} - \dfrac12\partial_x  \pi_{z}{}^{y} &= 0,\\
\dot D^{z}{}_{x} - \dfrac12 \partial_x\pi_{z}{}^{x} - \dfrac12\partial_x \pi_{x}{}^{z} &= 0, \\
 \dot D^{z}{}_{y} - \dfrac12 \partial_x\pi_{z}{}^{y} - \dfrac12 \partial_x\pi_{y}{}^{z} &= 0,\\
    \dot D^{z}{}_{z}   + \dfrac12 \partial_x \pi_{x}{}^{x}  + \dfrac12\partial_x \pi_{y}{}^{y} - \dfrac12\partial_x \pi_{z}{}^{z} & = 0,\\
    \dot D^{0}{}_{x} + \partial_x( D^{x}{}_{x} + D^{y}{}_{y} + D^{z}{}_{z}) & = 0,\\
    \dot D^{0}{}_{y} & = 0, \\
    \dot D^{0}{}_{z} & = 0,
    \end{split}
\end{equation}
and for the momenta in terms of the new variables;
\begin{equation}
\begin{split}
 \dot{\pi}_{y}{}^{y}   -\partial_x D^{z}{}_{z} & = 0,\\ 
     \dot{\pi}_{y}{}^{z} + \dfrac{1}{2} \partial_x D^{y}{}_{z} + \dfrac{1}{2} \partial_x D^{z}{}_{y} & = 0,\\
     \dot{\pi}_{z}{}^{y} + \dfrac{1}{2} \partial_x D^{z}{}_{y} + \dfrac{1}{2} \partial_x D^{y}{}_{z} & = 0, \\
    \dot{\pi}_{z}{}^{z}   -  \partial_x D^{y}{}_{y} & = 0,\\
    \dot{\pi}_0{}^x + \partial_x \pi_{x}{}^x & = 0,\\
\dot{\pi}_0{}^y + \partial_x \pi_{y}{}^{x} - \dfrac12 \partial_x {D^0}_y & = 0,\\
\dot{\pi}_0{}^z + \partial_x \pi_{z}{}^{x} - \dfrac12 \partial_x {D^0}_y & = 0.
    \label{pidoteqs_rew}
\end{split}
\end{equation}

\subsection{Properties of the 1-Dimensional Principal Symbol}
In order to extract the principal symbol from the first-order system presented in Eq.~\eqref{thetadoteqs_rew}-\eqref{pidoteqs_rew}, we define the vector $u_i$ of 24 components by collecting the momenta and auxiliary variables as 
\begin{equation}
u  =  ( \pi_{x}{}^{i},\pi_{y}{}^{i},\pi_{z}{}^{i},\pi_{0}{}^{i},D^{x}{}_{i},D^{y}{}_{i},D^{z}{}_{i},D^{0}{}_{i} ),
\end{equation}
with $i=x,y,z$.
We note that, unlike GR, the lapse function is not included since its derivatives do not appear in the equations. This departure from GR is justified since the boundary terms in the Hamiltonian are different. In our case, we consider the system in Eq.~\eqref{diffeqgeneral} with $q_i = 0$. After careful inspection of the system in Eq.~\eqref{thetadoteqs_rew} and Eq.~\eqref{pidoteqs_rew}, we acquire the principal symbol in Eq.~\eqref{principal1d} with the following eigenvalues
\begin{equation*}
\begin{split}
    &\lambda_1  = i, \ \ \ \qquad \text{with multiplicity}\quad2, \\ 
    &\lambda_2  = -i,  \qquad \text{ with multiplicity}\quad2, \\
    &\lambda_3  = 0,  \ \ \qquad \text{ with multiplicity}\quad20. 
\end{split}
\end{equation*}
and eigenvectors in Eq.~\eqref{vector1d}.

The presence of imaginary eigenvalues in the linearised one-dimensional reduction signals the loss of hyperbolicity at the level of the principal symbol. This result, although pessimistic, should be considered in the context of the assumptions that we had in mind: vanishing shift perturbations, gauge specialisation, and omission of Lagrange multipliers. The conclusion of loss of hyperbolicity \footnote{We assume that there is a strongly (symmetric) hyperbolic form at the nonlinear level that is lost under our simplifications.} applies only to the limited system in consideration and is not a statement about the predictability of TEGR, which has to be evaluated at the non-linear level. This evaluation is complicated, and partial computations pointing towards its hyperbolicity can be found in \cite{Peshkov:2022cbi}. 

\subsection{Modified 1-Dimensional System}
In the 1-dimensional analysis with restricted dependence on the $x$-coordinate, we can introduce new variables that depend only on the $y$ and $z$-coordinates in the following way
\begin{equation}
    \begin{split}
        F&:={\pi_y}^z+{\pi_z}^y,\\
        G&:={\pi_y}^y+{\pi_z}^z,\\
        H&:={\pi_y}^y-{\pi_z}^z,\\
        Q&:={D^y}_z+{D^z}_y,\\
        T&:={D^y}_y+{D^z}_z,\\
        R&:={D^y}_y-{D^z}_z.
    \end{split}
\end{equation}
Correspondingly, we have the modified system of evolution equations as follows,
\begin{equation}
    \begin{split}
        &\dot D\indices{^x_x}-\frac{1}{2}\partial_x{\pi_x}^x+\frac{1}{2}\partial_xG=0,\\
        &\dot D\indices{^x_y}-\frac{1}{2}\partial_x{\pi_x}^y-\frac{1}{2}\partial_x{\pi_y}^x=0,\\
        &\dot D\indices{^x_z}-\frac{1}{2}\partial_x{\pi_x}^z-\frac{1}{2}\partial_x{\pi_z}^x=0,\\
        &\dot D\indices{^y_x}-\frac{1}{2}\partial_x{\pi_x}^y--\frac{1}{2}\partial_x{\pi_y}^x=0,\\
        &\dot D\indices{^z_x}-\frac{1}{2}\partial_x{\pi_x}^z-\frac{1}{2}\partial_x{\pi_z}^x=0,\\
        &\dot D\indices{^0_x}+\partial_x{D^x}_x+\partial_xT=0,
    \end{split}
\end{equation}
\begin{equation}
    \begin{split}
        &\dot\pi\indices{_0^x}+\partial_x{\pi_x}^x=0,\\
        &\dot\pi\indices{_0^y}+\partial_x{\pi_y}^x-\frac{1}{2}\partial_x{D^0}_y=0,\\
        &\dot\pi\indices{_0^z}+\partial_x{\pi_z}^x-\frac{1}{2}\partial_x{D^0}_z=0,
    \end{split}
\end{equation}
\begin{equation}
    \begin{split}
        \dot F+\partial_xQ&=0,\\
        \dot Q-\partial_xF&=0,\\
        \dot H+\partial_xR&=0,\\
        \dot R-\partial_xH&=0,\\
        \dot G-\partial_xT&=0,\\
        \dot T=\dot D\indices{^0_y}=\dot D\indices{^0_z}=\dot{\pi}\indices{_x^y}&=\dot{\pi}\indices{_x^z}=\dot{\pi}\indices{_z^x}=\dot{\pi}\indices{_y^x}=0,\\
    \end{split}
\end{equation}

We will also apply the linearised constraints:
\begin{equation}
    \begin{split}
\mathcal{H}_x & = \partial_x \pi_{x}{}^{x} = 0,\\
\mathcal{H}_y & = \partial_x \pi_{y}{}^{x} = 0,\\
\mathcal{H}_z & = \partial_x \pi_{z}{}^{x} = 0,\\
C^{x} & = \pi_{0}{}^{x}  -T = 0\\
C^{y} & = \pi_{0}{}^{y} + D^{x}{}_{y} = 0,\\
C^{z} & = \pi_{0}{}^{z} + D^{x}{}_{z} = 0,\\
C^{xy} & = - C^{yx} = \pi_{y}{}^{x} - \pi_{x}{}^{y}  + 2 {D^{0}}_y = 0, \\
C^{xz} & = - C^{zx} =  \pi_{z}{}^{x} - \pi_{x}{}^{z}  + 2 {D^0{}}_{z}  = 0, \\
  C^{yz} & = - C^{zy} = \pi_{z}{}^{y} - \pi_{y}{}^{z}  = 0,\\
  \mathcal{H}_0 & = \partial_x \pi_0{}^{x} = 0.
\end{split}
\end{equation}
The modified system would only include three decoupled subsystem
\begin{equation}
    \begin{split}
        \dot D\indices{^x_x}+\frac{1}{2}\partial_xG=0,\\
        \dot D\indices{^0_x}+\partial_x{D^x}_x=0,\\
        \dot G=0,
    \end{split}
\end{equation}
where we can define
\begin{equation}
    C:= G+D\indices{^0_x}
\end{equation}
such that
\begin{equation}
\begin{split}
    \dot D\indices{^x_x}+\frac{1}{2}\partial_xC=0,\\
        \dot C+\partial_x{D^x}_x=0,
\end{split}
\end{equation}
\begin{equation}
    \begin{split}
        \dot F+\partial_xQ&=0,\\
        \dot Q-\partial_xF&=0,
    \end{split}
\end{equation}
\begin{equation}
    \begin{split}
        \dot H+\partial_xR&=0,\\
        \dot R-\partial_xH&=0.
    \end{split}
\end{equation}
\
We will also redefine the vector $u_i$
\begin{equation}
    u_i:=(u_1,u_2,u_2),
\end{equation}
with
\begin{equation}
    u_1:=({D^x}_x,C)\quad u_2:=(F,Q)\quad u_3:=(H,R).
\end{equation}
Correspondingly, the sub-principal matrices are
\begin{equation}
    M_1=\begin{pmatrix}
    0&\frac{1}{2}\\
    1&0\\
    \end{pmatrix},
\end{equation}
with eigenvalues $\lambda_1=1/\sqrt2,\ \lambda_2=-1/\sqrt{2}$, and
\begin{equation}
    M_2=M_3=\begin{pmatrix}
    0&1\\
    -1&0
    \end{pmatrix}.
\end{equation}
with eigenvalues $\lambda=\pm i$. These two decoupled subsystem with $u_2$ and $u_3$ form the rotational pairs that are responsible for the imaginary eigenvalues in the original system, and we can fix the gauge for them to initially be zero, therefore getting rid of them in the full system and of their ill-posed evolution. On the other hand, the modified main system is now in a strongly hyperbolic form, and it consists of a pair of variables $H,R$ that represent a well-posed perturbation mode. In this work, we are interested in proving that TEGR linear equations can be recast as a strongly hyperbolic system. We do not prove it, but the physical interpretation of the only propagating mode should correspond to one of the two polarization modes of gravitational waves in GR. We do not see the expected remaining propagating mode, most likely due to intensive gauge fixing, but it should appear with a different treatment of the system of differential equations.

In relation to our first result without redefinition of variables and using constraints, it has been reported in previous works in general relativity in the tetrad formalism that the principal part of the equations does not appear to be of any standard type, that is, neither hyperbolic nor parabolic \cite{Garfinkle:2005ph}. This leads to the lack of well-posedness for tetrad general relativity, but the system can be brought to be well-posed by adding constraints to the evolution equations. This connects with our previous result where the system was not hyperbolic, and supports the argument that it is not uncommon to find problems achieving well-posedness in tetrad-based gravities before working out constraints and redefinitions. 

Our toy model considered was originally designed to expose pathologies in the principal symbol with less effort. In the following, we explore whether some of the pathologies appearing in the one-dimensional case disappear or persist once we include dependence on all spatial dimensions.

\subsection{Properties of the principal symbol in 3D}

We now allow perturbations to depend on all three spatial coordinates $(x,y,z)$. Our goal is to assess the influence of the extra dimensions in the hyperbolicity behaviour. In particular, to diagnose whether the non-hyperbolic behaviour found in the raw one-dimensional case (without redefinitions of variables and constraints) is an artifact of the dimensional reduction or a genuine feature of the chosen Hamiltonian formulation and constraint structure.

We observe that the evolution equations for the tetrad components remain the same as in Eq.~\eqref{dhdt_diag} and Eq.~\eqref{dhdt_offdiag}. They are included in the full system analogously to the 1D case, that is, we compute the spatial derivatives $\partial_x, \partial_y$ and $\partial_z$ in all the equations, and perform the first-order reduction there. Additionally, the time evolution of the temporal part of the tetrad is also included, now with some modifications
\begin{equation}
    \begin{split}
       & \dot{h}^{0}{}_{x} + \partial_{x}( h^{x}{}_{x} + h^{y}{}_{y} + h^{z}{}_{z} ) = 0, \\
        & \dot{h}^{0}{}_{y} + \partial_{y}( h^{x}{}_{x} + h^{y}{}_{y} + h^{z}{}_{z} ) = 0,  \\
        & \dot{h}^{0}{}_{z} + \partial_{z}( h^{x}{}_{x} + h^{y}{}_{y} + h^{z}{}_{z} ) = 0.
    \end{split}
\end{equation}

The changes of the momenta evolution equations are less trivial, but can be summarized as
\begin{equation}
    \begin{split}
        - \dot\pi_{x}{}^{x} = & - \partial_{y} \partial_{y} h^{z}{}_{z} + \partial_{z} \partial_{y} h^{y}{}_{z} + \partial_{y} \partial_{z} h^{z}{}_{y} - \partial_{z} \partial_{z} h^{y}{}_{y}, \\
          - \dot\pi_{x}{}^{y} = & \dfrac12 \partial_{z} \partial_{z} ( h^{x}{}_{y} + h^{y}{}_{x} ) - \dfrac12  \partial_{z} \partial_{y} ( h^{x}{}_{z} +  h^{z}{}_{x})\\& - \dfrac12 \partial_{z} \partial_{x} ( h^{y}{}_{z} + h^{z}{}_{y} ) + \partial_{y} \partial_{x} h^{z}{}_{z}, \\
        - \dot\pi_{x}{}^{z} = & \dfrac12\partial_{y} \partial_{y} (h^{x}{}_{z} + h^{z}{}_{x}  ) - \dfrac12 \partial_{z} \partial_{y} (h^{x}{}_{y} + h^{y}{}_{x} )\\  & - \dfrac12 \partial_{y}\partial_{x} (h^{y}{}_{z} + h^{z}{}_{y}) + \partial_{z} \partial_{x} h^{y}{}_{y},\\
        - \dot\pi_{y}{}^{x} = &\dfrac12 \partial_{z} \partial_{z} ( h^{x}{}_{y} + h^{y}{}_{x} ) - \dfrac12 \partial_{z} \partial_{y} ( h^{x}{}_{z} +  h^{z}{}_{x})\\ & - \dfrac12 \partial_{z} \partial_{x} ( h^{y}{}_{z} + h^{z}{}_{y} ) + \partial_{y} \partial_{x} h^{z}{}_{z}, \\
        - \dot\pi_{y}{}^{y} =& -\partial_{z} \partial_{z} h^{x}{}_{x} - \partial_{x} \partial_{x} h^{z}{}_{z} + \partial_{z} \partial_{x} ( h^{x}{}_{z} + h^{z}{}_{x} ),\\
        - \dot\pi_{y}{}^{z} =& \dfrac12 \partial_{x} \partial_{x} ( h^{y}{}_{z} + h^{z}{}_{y} ) - \dfrac12 \partial_{z} \partial_{x} ( h^{x}{}_{y} + h^{y}{}_{x} ) \\&- \dfrac12 \partial_{y} \partial_{x} ( h^{x}{}_{z} + h^{z}{}_{x} ) + \partial_{z} \partial_{y} h^{x}{}_{x}, \\
        - \dot\pi_{z}{}^{x} =&  \dfrac12\partial_{y} \partial_{y} (h^{x}{}_{z} + h^{z}{}_{x}  ) - \dfrac12 \partial_{z} \partial_{y} (h^{x}{}_{y} + h^{y}{}_{x} )  \\&- \dfrac12 \partial_{y}\partial_{x} (h^{y}{}_{z} + h^{z}{}_{y} ) + \partial_{z} \partial_{x} h^{y}{}_{y},\\
        - \dot\pi_{z}{}^{y} =& \dfrac12 \partial_{x} \partial_{x} ( h^{y}{}_{z} + h^{z}{}_{x} ) - \dfrac12 \partial_{z} \partial_{x} ( h^{x}{}_{y} + h^{y}{}_{x} ) \\&- \dfrac12 \partial_{y} \partial_{x} ( h^{x}{}_{z} + h^{z}{}_{x} ) + \partial_{z} \partial_{y} h^{x}{}_{x},  \\
        - \dot\pi_{z}{}^{z} =& -\partial_{y} \partial_{y} h^{x}{}_{x} - \partial_{x} \partial_{x} h^{y}{}_{y} + \partial_{y} \partial_{x} ( h^{x}{}_{y} + h^{y}{}_{x} ).
    \end{split}
    \label{3dmomentaevolution}
\end{equation}
The temporal part of the momenta also have the following evolution equations
\begin{equation}
\begin{split}
    -\dot{\pi}_0{}^{x} & = \partial_x \pi_{x}{}^{x} + \partial_{y} \pi_{x}{}^{y} + \partial_{z} \pi_{x}{}^{z} \\
    & + \dfrac12( \partial_x(\partial_y h^{0}{}_{y} + \partial_{z} h^{0}{}_{z} ) - \partial_y \partial_y h^{0}{}_{x} - \partial_{z}\partial_{z} h^{0}{}_{x} ), \\
    -\dot{\pi}_0{}^{y} & = \partial_{x} \pi_{y}{}^{x} + \partial_{y} \pi_{y}{}^{y} + \partial_{z}\pi_{y}{}^{z} \\
    & + \dfrac12 ( \partial_{y}(\partial_{x} h^{0}{}_{x} + \partial_{z} h^{0}{}_{z} ) - \partial_{x} \partial_{x} h^{0}{}_{y} - \partial_{z} \partial_{z} h^{0}{}_{y} ), \\
    -\dot{\pi}_0{}^{z} & = \partial_{x} \pi_{z}{}^{x} + \partial_{y} \pi_{z}{}^{y} + \partial_{z}\pi_{z}{}^{z} \\
    & + \dfrac12 ( \partial_{z}(\partial_{x} h^{0}{}_{x} + \partial_{y}h^{0}{}_{y} ) - \partial_{x} \partial_{x} h^{0}{}_{z} - \partial_{y} \partial_{y} h^{0}{}_{z} ).
\end{split}
\end{equation}
Similar to the 1D case, we can define the spatial derivative of the metric perturbations as 
\begin{equation}
    {{D_x}^\mu}_j:=\partial_x{h^\mu}_j,\quad    
    {{D_y}^\mu}_j:=\partial_y{h^\mu}_j,\quad    
    {{D_z}^\mu}_j:=\partial_z{h^\mu}_j.
\end{equation}
The linearised constraints in Eq.~\eqref{LorC1} with full dependence on three spatial coordinates are
\begin{equation}
    \begin{split}
        C^x & = -\pi_0{}^x - \partial_x ( h^{y}{}_{y} + h^{z}{}_{z} ) + \partial_y h^{y}{}_{x} + \partial_z h^{z}{}_{x}, \\
        C^{y} & = -\pi_0{}^y + \partial_{x} h^{x}{}_{y} - \partial_y ( h^{x}{}_{x} + h^{z}{}_{z} ) + \partial_{z} h^{z}{}_{y}, \\
        C^{z} & = -\pi_0{}^z + \partial_{x} h^{x}{}_{z} + \partial_{y} h^{y}{}_{z} - \partial_{z}( h^{x}{}_{x} + h^{y}{}_{y} ).
    \end{split}
\end{equation}
Meanwhile, the linearized constraints \eqref{LinLorentz2} are generalised as 
\begin{equation}
\begin{split}
   C^{xy} & = - C^{yx} = \dfrac12 ( \pi_{y}{}^{x} - \pi_{x}{}^{y} ) + \partial_x h^{0}{}_{y} - \partial_y h^{0}{}_{x}, \\
   C^{xz} & = - C^{zx} = \dfrac12 ( \pi_{z}{}^{x} - \pi_{x}{}^{z} ) + \partial_x h^{0}{}_{z} - \partial_z h^{0}{}_{x}, \\
   C^{yz} & = - C^{zy} = \dfrac12 ( \pi_{z}{}^{y} - \pi_{y}{}^{z} ) + \partial_y h^{0}{}_{z} - \partial_z h^{0}{}_{y},
\end{split}
\end{equation}
when compared to \eqref{Lor2lin}. Finally, the Hamiltonian constraint has a more general form
\begin{equation}
\mathcal{H}_0 = \partial_x \pi_0{}^x + \partial_y \pi_0{}^y + \partial_z \pi_0{}^z = 0.
\end{equation}
The vector that defines the dynamical variables of the system of differential equations includes the conjugate momenta to the spatial tetrad, and the partial derivatives of the spatial tetrad with respect to all three spatial dimensions, and is defined as follows:
\begin{equation}
    u_i:=({\pi_\mu}^i, {{D_x}^\mu}_j,{{D_y}^\mu}_j,{{D_z}^\mu}_j)
\end{equation}
where $a=\{0,i\}$ and $i=\{x,y,z\}$.\\
The three-dimensional first-order system of evolution equations for the set of variables $u_i$ is
\begin{equation}
    \partial_tu_i+M_{ij}^x\partial_xu_j+M_{ij}^y\partial_yu_j+M_{ij}^z\partial_zu_j=q_i,
\end{equation}
where we take $q_i=0$.\\
For the system to be strongly-hyperbolic, the principal symbol, $\mathbb{M}$, defined as 
\begin{equation}
    \mathbb{M}:= k_xM^x+k_yM^y+k_zM^z=\mathbf{kM},
\end{equation}
must be diagonalisable with real eigenvalues and have eigenvectors that span the whole space for any real values of $k_x$, $k_y$, and $k_z$.

Unlike the 1D system, the evolution of the conjugate momenta involves second order spatial derivatives in multiple spatial directions. Since the spatial derivatives commute, i.e $\partial_xD\indices{_y^\mu_j}=\partial_yD\indices{_x^\mu_j}$, there is a degeneracy in defining such a term in $M^x$ or $M^y$. However, this degeneracy does not affect the hyperbolicity property of the system. On the level of constraint-free system, they can be considered symbol-equivalent representations \cite{KreissLorenz2004,Hormander1985AnalysisIII}. Such equivalence preserves the hyperbolicity property of the evolution system. We will present a choice of the principal symbol by prioritising terms in $M^x$ over $M^y$ over $M^z$ at the lowest order in Appendix \ref{principal3d}.\\
The resulting principal symbol has the following eigenvalues
\begin{equation}
    \begin{split}
        &\lambda_1=0 \qquad\qquad\qquad\qquad\quad\ \text{ with multiplicity 44},\\
        &\lambda_2=i\sqrt{k_x^2+k_y^2+k_z^2}\quad\qquad\text{ with multiplicity 2},\\
        &\lambda_3=-i\sqrt{k_x^2+k_y^2+k_z^2}\ \qquad\text{ with multiplicity 2}.\\
    \end{split}
\end{equation}
Similar to the one-dimensional case, the system is clearly not hyperbolic as both $\lambda_2$ and $\lambda_3$ are purely imaginary.\\
The symbol-equivalent property that eliminates the ambiguity in choosing the order of taking spatial derivatives is only present when constraints are not applied to the system \cite{Gundlach:2006tw, Hormander1985AnalysisIII, KreissLorenz2004}. This is well known in the case of ADM, BSSN, and CCZ4 formalisms in GR \cite{Baumgarte:1998te, Shibata:1995we, Dumbser:2017okk}. In the case of additional constraints, the hyperbolicity property depends on the exact representation. In this system, the space of representations is so vast that it is very difficult to scan the entire space to pinpoint the set of representations and constraints that would lead to a hyperbolic or strongly hyperbolic system. Such a search is beyond the scope of this work and we will leave this for future studies.\\
\section{Conclusions and outlook}
\label{sec:conc}

We perform a principal-symbol analysis of the TEGR $3+1$ evolution equations derived from the covariant Hamiltonian formulation based on the VAST decomposition of the canonical variables. Although TEGR is dynamically equivalent to metric general relativity, its $3+1$ formulation may differ due to  boundary-term choices and to the manner in which gauge degrees of freedom enter the principal part of the evolution system. Moreover, tetrad-based formulations introduce additional variables and constraints absent in metric formulations. These additional structures can, in principle, be exploited to alter the properties of the principal-symbol through constraint additions and gauge choices, potentially affecting numerical stability and well-posedness.

We followed the ADM ``toy model'' approach in \cite{article} for GR, where it was shown that the ADM equations form a weakly hyperbolic system of differential equations at the linear level. Taking the Hamilton's equations for TEGR as a starting point, we linearised the canonical variables around Minkowski spacetime, specialised to vanishing shift, and performed a first-order reduction by introducing auxiliary variables for the spatial derivatives of tetrad perturbations. We then extract the principal symbol and utilised its eigenstructure to diagnose a potential (strong) hyperbolicity property.

The one-dimensional reduction (dependence on a single $x$ Cartesian coordinate) reveals that for the Hamiltonian evolution system with the Lagrange multipliers set to zero, the resulting principal symbol contains a sector with imaginary eigenvalues. Consequently, the system is not hyperbolic in this reduction. We further explored the freedom to redefine variables and add linear combinations of constraints to the evolution equations. Such interventions allowed us to isolate the ill-posed part of the evolution, and extract the set of variables that form a strongly hyperbolic system. 

Our original strategy for simplifying Hamilton's equation for the conjugate momenta \eqref{HeqPiFull} was to set all the Lagrange multipliers associated with primary constraints to zero. Although we have mentioned that they could be introduced back later to modify the properties of the system, it is worth noting that, at least for the form in which the constraints are currently expressed, it is not possible to obtain new spatial derivatives of the fields that would influence the principal symbol. All primary constraints are either linear in the conjugate variables or contain only first order derivatives of the tetrad perturbations. Since we wrote our system in terms of $D^{i}{}_{j}$, which already represent first-order derivatives, they do not appear with additional first-order spatial derivatives in the constraints.

We extended the analysis to the full three spatial dimensions by constructing the full principal symbol, $\mathbb M$. In the representation considered, the same non-hyperbolic feature appears for any $\mathbf{k} \neq 0$. This shows that the non-hyperbolicity observed in the one-dimensional reduction is not an artifact of the model or the dimensionality chosen. Due to the intricacy and extent of the system of equation, we leave attempts to prove the strong hyperbolicity of the 3D system for future work.

We note that it is not unusual for a given formulation of the dynamical equations in tetrad formalism to resist being put into a hyperbolic form. As an example, in \cite{Garfinkle:2005ph}, where the scale-invariant tetrad formulation of the vacuum Einstein equations introduced in \cite{Uggla:2003fp}, is reformulated as a symmetric hyperbolic/parabolic system. In \cite{Uggla:2003fp}, the motivation to use a set of scale invariant variables based on the tetrad formalism is to find a system suitable for the analysis of the properties of singularities and to study the asymptotic properties of the metric. Whether a similar trick can be made in TEGR at the linear level can be explored in future work. In contrast with more general torsion-based theories, in recent work in Einstein–Cartan theory, local geometric well‑posedness is established under a generalized harmonic gauge despite the presence of torsion \cite{Luz:2025rqx}, which could be used as a benchmark for non-linear studies of hyperbolicity in pure torsion theories.

The interpretation of the loss of hyperbolicity found in this work for the initial set of equations is not a statement about TEGR as a physical theory, but about a particular formulation at the linear order, which is a combination of choice of variables and gauge specialisation. There are many reasons why this system does not inherit straight away the well-posedness properties of strongly hyperbolic metric formulations, but we believe it might be related to the elimination of the Lagrange multipliers containing all the constraints, which are difficult to include. Therefore, an ill-behaved gauge propagation might remain and has superluminal propagation speeds. 

In the near future, we expect this work to facilitate the study of well-posedness in TEGR for physically relevant examples, such as spherical symmetric systems. The practical use of TEGR Hamilton's equations, together with the lessons learned in this work, should make this task feasible, particularly in regard to the careful formulation of the gauge sector and the possible inclusion of constraint-damping strategies. Since a strongly hyperbolic first-order system of differential equations is a necessity for running the first numerical relativity simulations in teleparallel gravity, the continuation of this work is essential and is currently under investigation.

\section*{Acknowledgments}

The authors are grateful to Miguel Bezares, Jos\'e Antonio Font, Thanasis Giannakopoulos, Marie Femke Jaarma, Guillermo Lara, Claudio Lazarte, Luis Lehner, Eugene Lim and Marcelo Rubio for several references pertinent to this work and many discussions that greatly contributed to forging the ideas presented here. Some of the computations in this paper have been assisted with Cadabra 2.0 \cite{Peeters:2007wn,Peeters:2018dyg,PEETERS2007550}. M.J.G. has been supported by the Estonian Research Council grant PSG910 ``Theoretical frameworks for numerical modified gravity''

\bibliographystyle{apsrev4-1}
\bibliography{main}
\appendix
\onecolumngrid
\clearpage
\section{1-Dimensional System}
\label{p1d}
The principal symbol of the 1-dimensional analysis is
\setcounter{MaxMatrixCols}{24}
\begin{equation}\label{principal1d}
    M_x =
\begin{pmatrix}
0&0&0&0&0&0&0&0&0&0&0&0&0&0&0&0&0&0&0&0&0&0&0&0\\
0&0&0&0&0&0&0&0&0&0&0&0&0&0&0&0&0&0&0&0&0&0&0&0\\
0&0&0&0&0&0&0&0&0&0&0&0&0&0&0&0&0&0&0&0&0&0&0&0\\
0&0&0&0&0&0&0&0&0&0&0&0&0&0&0&0&0&0&0&0&0&0&0&0\\
0&0&0&0&0&0&0&0&0&0&0&0&0&0&0&0&0&0&0&0&-1&0&0&0\\
0&0&0&0&0&0&0&0&0&0&0&0&0&0&0&0&0&\tfrac12&0&\tfrac12&0&0&0&0\\
0&0&0&0&0&0&0&0&0&0&0&0&0&0&0&0&0&0&0&0&0&0&0&0\\
0&0&0&0&0&0&0&0&0&0&0&0&0&0&0&0&0&\tfrac12&0&\tfrac12&0&0&0&0\\
0&0&0&0&0&0&0&0&0&0&0&0&0&0&0&0&-1&0&0&0&0&0&0&0\\
1&0&0&0&0&0&0&0&0&0&0&0&0&0&0&0&0&0&0&0&0&0&0&0\\
0&0&0&1&0&0&0&0&0&0&0&0&0&0&0&0&0&0&0&0&0&0&-\tfrac12&0\\
0&0&0&0&0&0&1&0&0&0&0&0&0&0&0&0&0&0&0&0&0&0&0&-\tfrac12\\
-\tfrac12&0&0&0&\tfrac12&0&0&0&\tfrac12&0&0&0&0&0&0&0&0&0&0&0&0&0&0&0\\
0&-\tfrac12&0&-\tfrac12&0&0&0&0&0&0&0&0&0&0&0&0&0&0&0&0&0&0&0&0\\
0&0&-\tfrac12&0&0&0&-\tfrac12&0&0&0&0&0&0&0&0&0&0&0&0&0&0&0&0&0\\
0&-\tfrac12&0&-\tfrac12&0&0&0&0&0&0&0&0&0&0&0&0&0&0&0&0&0&0&0&0\\
\tfrac12&0&0&0&-\tfrac12&0&0&0&\tfrac12&0&0&0&0&0&0&0&0&0&0&0&0&0&0&0\\
0&0&0&0&0&-\tfrac12&0&-\tfrac12&0&0&0&0&0&0&0&0&0&0&0&0&0&0&0&0\\
0&0&-\tfrac12&0&0&0&-\tfrac12&0&0&0&0&0&0&0&0&0&0&0&0&0&0&0&0&0\\
0&0&0&0&0&-\tfrac12&0&-\tfrac12&0&0&0&0&0&0&0&0&0&0&0&0&0&0&0&0\\
\tfrac12&0&0&0&\tfrac12&0&0&0&-\tfrac12&0&0&0&0&0&0&0&0&0&0&0&0&0&0&0\\
0&0&0&0&0&0&0&0&0&0&0&0&-1&0&0&0&-1&0&0&0&-1&0&0&0\\
0&0&0&0&0&0&0&0&0&0&0&0&0&0&0&0&0&0&0&0&0&0&0&0\\
0&0&0&0&0&0&0&0&0&0&0&0&0&0&0&0&0&0&0&0&0&0&0&0
\end{pmatrix}
\end{equation}
We denote the eigenvectors corresponding to the eigenvalues as $v_i^{(a)}$, where $i=\{1,2,...,24\}$ denotes the corresponding eigenvalue, and $a$ numbers the eigenvector corresponding to that eigenvalue, which are
\begin{equation}\label{vector1d}
\begin{split}
&v_1^i = (0,0,0,0,i,0,0,0,-i,0,0,0,0,0,0,0,-1,0,0,0,1,0,0,0),\\
&v_2^i = (0,0,0,0,0,-i,0,-i,0,0,0,0,0,0,0,0,0,1,0,1,0,0,0,0),\\
&v_3^{-i} = (0,0,0,0,-i,0,0,0,i,0,0,0,0,0,0,0,-1,0,0,0,1,0,0,0),\\
&v_4^{-i} = (0,0,0,0,0,i,0,i,0,0,0,0,0,0,0,0,0,1,0,1,0,0,0,0),\\
&v_5^0 = (0,0,-\tfrac{1}{2},0,0,0,\tfrac{1}{2},0,0,0,0,0,0,0,0,0,0,0,0,0,0,0,0,1),\\
&v_6^0 = (0,-\tfrac{1}{2},0,\tfrac{1}{2},0,0,0,0,0,0,0,0,0,0,0,0,0,0,0,0,0,0,1,0),\\
&v_7^0 = (0,0,0,0,0,0,0,0,0,0,0,0,0,0,0,0,0,0,0,0,0,1,0,0),\\
&v_8^0 = (0,0,0,0,0,0,0,0,0,0,0,0,0,0,0,0,0,-1,0,1,0,0,0,0),\\
&v_9^0 = (0,0,0,0,0,0,0,0,0,0,0,0,0,0,0,0,0,0,1,0,0,0,0,0),\\
&v_{10}^0 = (0,0,0,0,0,0,0,0,0,0,0,0,0,0,0,1,0,0,0,0,0,0,0,0),\\
&v_{11}^0 = (0,0,0,0,0,0,0,0,0,0,0,0,0,0,1,0,0,0,0,0,0,0,0,0),\\
&v_{12}^0 = (0,0,0,0,0,0,0,0,0,0,0,0,0,1,0,0,0,0,0,0,0,0,0,0),\\
&v_{13}^0 = (0,0,0,0,0,0,0,0,0,0,0,1,0,0,0,0,0,0,0,0,0,0,0,0),\\
&v_{14}^0 = (0,0,0,0,0,0,0,0,0,0,1,0,0,0,0,0,0,0,0,0,0,0,0,0),\\
&v_{15}^0 = (0,0,0,0,0,0,0,0,0,1,0,0,0,0,0,0,0,0,0,0,0,0,0,0),\\
&v_{16}^0 = (0,0,0,0,0,-1,0,1,0,0,0,0,0,0,0,0,0,0,0,0,0,0,0,0),\\
&v_{17-24}^0 = (0,0,0,0,0,0,0,0,0,0,0,0,0,0,0,0,0,0,0,0,0,0,0,0).
\end{split}
\end{equation}
\newpage
\twocolumngrid
\section{3-Dimensional System}
\label{principal3d}
For the convenience of the presentation, we split the principal symbol into 16 $12\times12$ matrix elements as follows
\begin{equation}
    \mathbb{M}:=\left(\begin{array}{cccc}
        M_{11} & M_{12}& M_{13}& M_{14}\\
        M_{21} & M_{22}& M_{23}& M_{24}\\
        M_{31} & M_{32}& M_{33}& M_{34}\\
        M_{41} & M_{42}& M_{43}& M_{44}\\
    \end{array}
\right).
\end{equation}\\
With our choice of representation, we have the following trivial elements in the principal symbol
\begin{equation}
    M_{32}=M_{42}=M_{43}=\textbf{0}_{12\times12}.
\end{equation}
The non-trivial matrix elements in the principal symbol are
\footnotesize
\begin{equation*}
    M_{11} =
\begin{pmatrix}
0&0&0&0&0&0&0&0&0&0&0&0\\
0&0&0&0&0&0&0&0&0&0&0&0\\
0&0&0&0&0&0&0&0&0&0&0&0\\
0&0&0&0&0&0&0&0&0&0&0&0\\
0&0&0&0&0&0&0&0&0&0&0&0\\
0&0&0&0&0&0&0&0&0&0&0&0\\
0&0&0&0&0&0&0&0&0&0&0&0\\
0&0&0&0&0&0&0&0&0&0&0&0\\
0&0&0&0&0&0&0&0&0&0&0&0\\
k_x&k_y&k_z&0&0&0&0&0&0&0&0&0\\
0&0&0&k_x&k_y&k_z&0&0&0&0&0&0\\
0&0&0&0&0&0&k_x&k_y&k_z&0&0&0
\end{pmatrix}
\end{equation*}\\
\begin{equation*}
    M_{12} =
\begin{pmatrix}
0&0&0&0&0&0&0&0&0&0&0&0\\
0&0&0&0&0&0&0&0&0&0&0&0\\
0&0&0&0&0&0&0&0&0&0&0&0\\
0&0&0&0&0&0&0&0&0&0&0&0\\
0&0&0&0&0&0&0&0&-k_x&0&0&0\\
0&0&0&0&0&\tfrac{k_x}{2}&0&\tfrac{k_x}{2}&0&0&0&0\\
0&0&0&0&0&0&0&0&0&0&0&0\\
0&0&0&0&0&\tfrac{k_x}{2}&0&\tfrac{k_x}{2}&0&0&0&0\\
0&0&0&0&-k_x&0&0&0&0&0&0&0\\
0&0&0&0&0&0&0&0&0&0&0&0\\
0&0&0&0&0&0&0&0&0&0&-\tfrac{k_x}{2}&0\\
0&0&0&0&0&0&0&0&0&0&0&-\tfrac{k_x}{2}
\end{pmatrix}
\end{equation*}\\
\scriptsize
\begin{equation*}
M_{13}=
\setlength{\arraycolsep}{2pt}
\begin{pmatrix}
0&0&0&0&0&0&0&0&-k_y&0&0&0\\
0&0&0&0&0&0&0&0&k_x&0&0&0\\
0&0&\tfrac{k_y}{2}&0&0&-\tfrac{k_x}{2}&\tfrac{k_y}{2}&-\tfrac{k_x}{2}&0&0&0&0\\
0&0&0&0&0&0&0&0&k_x&0&0&0\\
0&0&0&0&0&0&0&0&0&0&0&0\\
0&0&-\tfrac{k_x}{2}&0&0&0&-\tfrac{k_x}{2}&0&0&0&0&0\\
0&0&\tfrac{k_y}{2}&0&0&-\tfrac{k_x}{2}&\tfrac{k_y}{2}&-\tfrac{k_x}{2}&0&0&0&0\\
0&0&-\tfrac{k_x}{2}&0&0&0&-\tfrac{k_x}{2}&0&0&0&0&0\\
-k_y&k_x&0&k_x&0&0&0&0&0&0&0&0\\
0&0&0&0&0&0&0&0&0&-\tfrac{k_y}{2}&\tfrac{k_x}{2}&0\\
0&0&0&0&0&0&0&0&0&\tfrac{k_x}{2}&0&0\\
0&0&0&0&0&0&0&0&0&0&0&-\tfrac{k_y}{2}\\
\end{pmatrix}
\end{equation*}\\
\begin{equation*}
M_{14}=
\setlength{\arraycolsep}{1pt}
\begin{pmatrix}
0&0&0&0&-k_z&k_y&0&k_y&0&0&0&0\\
0&\tfrac{k_z}{2}&-\tfrac{k_y}{2}&\tfrac{k_z}{2}&0&-\tfrac{k_x}{2}&-\tfrac{k_y}{2}&-\tfrac{k_x}{2}&0&0&0&0\\
0&-\tfrac{k_y}{2}&0&-\tfrac{k_y}{2}&k_x&0&0&0&0&0&0&0\\
0&\tfrac{k_z}{2}&-\tfrac{k_y}{2}&\tfrac{k_z}{2}&0&-\tfrac{k_x}{2}&-\tfrac{k_y}{2}&-\tfrac{k_x}{2}&0&0&0&0\\
-k_z&0&k_x&0&0&0&k_x&0&0&0&0&0\\
k_y&-\tfrac{k_x}{2}&0&-\tfrac{k_x}{2}&0&0&0&0&0&0&0&0\\
0&-\tfrac{k_y}{2}&0&-\tfrac{k_y}{2}&k_x&0&0&0&0&0&0&0\\
k_y&-\tfrac{k_x}{2}&0&-\tfrac{k_x}{2}&0&0&0&0&0&0&0&0\\
0&0&0&0&0&0&0&0&0&0&0&0\\
0&0&0&0&0&0&0&0&0&-\tfrac{k_z}{2}&0&-\tfrac{k_x}{2}\\
0&0&0&0&0&0&0&0&0&0&-\tfrac{k_z}{2}&\tfrac{k_y}{2}\\
0&0&0&0&0&0&0&0&0&\tfrac{k_x}{2}&\tfrac{k_y}{2}&0\\
\end{pmatrix}
\end{equation*}\\
\footnotesize
\begin{equation*}
    M_{21} =k_x
\begin{pmatrix}
-\tfrac{1}{2}&0&0&0&\tfrac{1}{2}&0&0&0&\tfrac{1}{2}&0&0&0\\
0&-\tfrac{1}{2}&0&-\tfrac{1}{2}&0&0&0&0&0&0&0&0\\
0&0&-\tfrac{1}{2}&0&0&0&-\tfrac{1}{2}&0&0&0&0&0\\
0&-\tfrac{1}{2}&0&-\tfrac{1}{2}&0&0&0&0&0&0&0&0\\
\tfrac{1}{2}&0&0&0&-\tfrac{1}{2}&0&0&0&\tfrac{1}{2}&0&0&0\\
0&0&0&0&0&-\tfrac{1}{2}&0&-\tfrac{1}{2}&0&0&0&0\\
0&0&-\tfrac{1}{2}&0&0&0&-\tfrac{1}{2}&0&0&0&0&0\\
0&0&0&0&0&-\tfrac{1}{2}&0&-\tfrac{1}{2}&0&0&0&0\\
\tfrac{1}{2}&0&0&0&\tfrac{1}{2}&0&0&0&-\tfrac{1}{2}&0&0&0\\
0&0&0&0&0&0&0&0&0&0&0&0\\
0&0&0&0&0&0&0&0&0&0&0&0\\
0&0&0&0&0&0&0&0&0&0&0&0
\end{pmatrix}
\end{equation*}\\
\begin{equation*}
    M_{22} =
\begin{pmatrix}
0&0&0&0&0&0&0&0&0&0&0&0\\
0&0&0&0&0&0&0&0&0&0&0&0\\
0&0&0&0&0&0&0&0&0&0&0&0\\
0&0&0&0&0&0&0&0&0&0&0&0\\
0&0&0&0&0&0&0&0&0&0&0&0\\
0&0&0&0&0&0&0&0&0&0&0&0\\
0&0&0&0&0&0&0&0&0&0&0&0\\
0&0&0&0&0&0&0&0&0&0&0&0\\
0&0&0&0&0&0&0&0&0&0&0&0\\
k_x&0&0&0&k_x&0&0&0&k_x&0&0&0\\
0&0&0&0&0&0&0&0&0&0&0&0\\
0&0&0&0&0&0&0&0&0&0&0&0
\end{pmatrix}
\end{equation*}\\
\begin{equation*}
    M_{23} =
\begin{pmatrix}
0&0&0&0&0&0&0&0&0&0&0&0\\
0&0&0&0&0&0&0&0&0&0&0&0\\
0&0&0&0&0&0&0&0&0&0&0&0\\
0&0&0&0&0&0&0&0&0&0&0&0\\
0&0&0&0&0&0&0&0&0&0&0&0\\
0&0&0&0&0&0&0&0&0&0&0&0\\
0&0&0&0&0&0&0&0&0&0&0&0\\
0&0&0&0&0&0&0&0&0&0&0&0\\
0&0&0&0&0&0&0&0&0&0&0&0\\
0&0&0&0&0&0&0&0&0&0&0&0\\
k_x&0&0&0&k_x&0&0&0&k_x&0&0&0\\
0&0&0&0&0&0&0&0&0&0&0&0
\end{pmatrix}
\end{equation*}\\
\begin{equation*}
    M_{24} =
\begin{pmatrix}
0&0&0&0&0&0&0&0&0&0&0&0\\
0&0&0&0&0&0&0&0&0&0&0&0\\
0&0&0&0&0&0&0&0&0&0&0&0\\
0&0&0&0&0&0&0&0&0&0&0&0\\
0&0&0&0&0&0&0&0&0&0&0&0\\
0&0&0&0&0&0&0&0&0&0&0&0\\
0&0&0&0&0&0&0&0&0&0&0&0\\
0&0&0&0&0&0&0&0&0&0&0&0\\
0&0&0&0&0&0&0&0&0&0&0&0\\
0&0&0&0&0&0&0&0&0&0&0&0\\
0&0&0&0&0&0&0&0&0&0&0&0\\
k_x&0&0&0&k_x&0&0&0&k_x&0&0&0
\end{pmatrix}
\end{equation*}\\
\begin{equation*}
    M_{31} =k_y
\begin{pmatrix}
-\tfrac{1}{2}&0&0&0&\tfrac{1}{2}&0&0&0&\tfrac{1}{2}&0&0&0\\
0&-\tfrac{1}{2}&0&-\tfrac{1}{2}&0&0&0&0&0&0&0&0\\
0&0&-\tfrac{1}{2}&0&0&0&-\tfrac{1}{2}&0&0&0&0&0\\
0&-\tfrac{1}{2}&0&-\tfrac{1}{2}&0&0&0&0&0&0&0&0\\
\tfrac{1}{2}&0&0&0&-\tfrac{1}{2}&0&0&0&\tfrac{1}{2}&0&0&0\\
0&0&0&0&0&-\tfrac{1}{2}&0&-\tfrac{1}{2}&0&0&0&0\\
0&0&-\tfrac{1}{2}&0&0&0&-\tfrac{1}{2}&0&0&0&0&0\\
0&0&0&0&0&-\tfrac{1}{2}&0&-\tfrac{1}{2}&0&0&0&0\\
\tfrac{1}{2}&0&0&0&\tfrac{1}{2}&0&0&0&-\tfrac{1}{2}&0&0&0\\
0&0&0&0&0&0&0&0&0&0&0&0\\
0&0&0&0&0&0&0&0&0&0&0&0\\
0&0&0&0&0&0&0&0&0&0&0&0
\end{pmatrix}
\end{equation*}\\
\clearpage
\onecolumngrid
\noindent
\begin{minipage}[t]{0.48\textwidth}
\begin{equation*}
    M_{33} =
\begin{pmatrix}
0&0&0&0&0&0&0&0&0&0&0&0\\
0&0&0&0&0&0&0&0&0&0&0&0\\
0&0&0&0&0&0&0&0&0&0&0&0\\
0&0&0&0&0&0&0&0&0&0&0&0\\
0&0&0&0&0&0&0&0&0&0&0&0\\
0&0&0&0&0&0&0&0&0&0&0&0\\
0&0&0&0&0&0&0&0&0&0&0&0\\
0&0&0&0&0&0&0&0&0&0&0&0\\
0&0&0&0&0&0&0&0&0&0&0&0\\
k_x&0&0&0&k_x&0&0&0&k_x&0&0&0\\
k_y&0&0&0&k_y&0&0&0&k_y&0&0&0\\
0&0&0&0&0&0&0&0&0&0&0&0
\end{pmatrix}
\end{equation*}
\vspace{8mm}

\begin{equation*}
    M_{34} =
\begin{pmatrix}
0&0&0&0&0&0&0&0&0&0&0&0\\
0&0&0&0&0&0&0&0&0&0&0&0\\
0&0&0&0&0&0&0&0&0&0&0&0\\
0&0&0&0&0&0&0&0&0&0&0&0\\
0&0&0&0&0&0&0&0&0&0&0&0\\
0&0&0&0&0&0&0&0&0&0&0&0\\
0&0&0&0&0&0&0&0&0&0&0&0\\
0&0&0&0&0&0&0&0&0&0&0&0\\
0&0&0&0&0&0&0&0&0&0&0&0\\
0&0&0&0&0&0&0&0&0&0&0&0\\
0&0&0&0&0&0&0&0&0&0&0&0\\
k_y&0&0&0&k_y&0&0&0&k_y&0&0&0
\end{pmatrix}
\end{equation*}
\end{minipage}
\hfill
\begin{minipage}[t]{0.48\textwidth}
\vspace{-2mm}
\begin{equation*}
    M_{41} =k_z
\begin{pmatrix}
-\tfrac{1}{2}&0&0&0&\tfrac{1}{2}&0&0&0&\tfrac{1}{2}&0&0&0\\
0&-\tfrac{1}{2}&0&-\tfrac{1}{2}&0&0&0&0&0&0&0&0\\
0&0&-\tfrac{1}{2}&0&0&0&-\tfrac{1}{2}&0&0&0&0&0\\
0&-\tfrac{1}{2}&0&-\tfrac{1}{2}&0&0&0&0&0&0&0&0\\
\tfrac{1}{2}&0&0&0&-\tfrac{1}{2}&0&0&0&\tfrac{1}{2}&0&0&0\\
0&0&0&0&0&-\tfrac{1}{2}&0&-\tfrac{1}{2}&0&0&0&0\\
0&0&-\tfrac{1}{2}&0&0&0&-\tfrac{1}{2}&0&0&0&0&0\\
0&0&0&0&0&-\tfrac{1}{2}&0&-\tfrac{1}{2}&0&0&0&0\\
\tfrac{1}{2}&0&0&0&\tfrac{1}{2}&0&0&0&-\tfrac{1}{2}&0&0&0\\
0&0&0&0&0&0&0&0&0&0&0&0\\
0&0&0&0&0&0&0&0&0&0&0&0\\
0&0&0&0&0&0&0&0&0&0&0&0
\end{pmatrix}
\end{equation*}
\vspace{7mm}
\begin{equation*}
    M_{44} =
\begin{pmatrix}
0&0&0&0&0&0&0&0&0&0&0&0\\
0&0&0&0&0&0&0&0&0&0&0&0\\
0&0&0&0&0&0&0&0&0&0&0&0\\
0&0&0&0&0&0&0&0&0&0&0&0\\
0&0&0&0&0&0&0&0&0&0&0&0\\
0&0&0&0&0&0&0&0&0&0&0&0\\
0&0&0&0&0&0&0&0&0&0&0&0\\
0&0&0&0&0&0&0&0&0&0&0&0\\
0&0&0&0&0&0&0&0&0&0&0&0\\
k_x&0&0&0&k_x&0&0&0&k_x&0&0&0\\
k_y&0&0&0&k_y&0&0&0&k_y&0&0&0\\
k_z&0&0&0&k_z&0&0&0&k_z&0&0&0
\end{pmatrix}
\end{equation*}
\end{minipage}

\end{document}